\documentclass[aps,prb,preprint,superscriptaddress]{revtex4-1}
\usepackage{graphicx,amsmath,amsfonts,mathrsfs}

\begin{document}

\title{Validity and limitations of the superexchange model for the
magnetic properties of Sr$_2$IrO$_4$ and Ba$_2$IrO$_4$ mediated by the strong spin-orbit coupling}
\author{I. V. Solovyev}
\email{SOLOVYEV.Igor@nims.go.jp}
\affiliation{Computational Materials Science Unit, National Institute for Materials
Science, 1-1 Namiki, Tsukuba, Ibaraki 305-0044, Japan}
\affiliation{Department of Theoretical Physics and Applied Mathematics, Ural Federal
University, Mira str. 19, 620002 Ekaterinburg, Russia}
\author{V. V. Mazurenko}
\affiliation{Department of Theoretical Physics and Applied Mathematics, Ural Federal
University, Mira str. 19, 620002 Ekaterinburg, Russia}
\author{A. A. Katanin}
\affiliation{Institute of Metal Physics, S. Kovalevskoy Str. 18, 620990 Ekaterinburg,
Russia}
\affiliation{Department of Theoretical Physics and Applied Mathematics, Ural Federal
University, Mira str. 19, 620002 Ekaterinburg, Russia}
\date{\today}

\begin{abstract}
Layered perovskites Sr$_2$IrO$_4$ and Ba$_2$IrO$_4$ are regarded as the key materials for understanding the
properties of magnetic relativistic insulators, mediated by the strong spin-orbit (SO) coupling.
One of the most fundamental issues is
to which extent these properties can be
described by the superexchange (SE) model, formulated in the limit of the large Coulomb repulsion for
some appropriately selected pseudospin states, and whether these materials themselves
can be classified as Mott insulators. In the present work we address these issues by deriving the
relevant models and extracting parameters of these models from the first-principles electronic structure calculations
with the SO coupling.
First, we construct the effective Hubbard-type model for the magnetically active $t_{2g}$ bands,
by recasting the problem in the language of localized Wannier orbitals.
Then, we map the so obtained electron model onto the pseudospin model by applying the theory
of SE interactions, which is based on the second-order perturbation theory with respect to the
transfer integrals. We discuss the microscopic origin of anisotropic SE interactions,
inherent to the compass Heisenberg model, and the appearance of
the antisymmetric Dzyaloshinskii-Moriya term, associated with the additional rotation of the IrO$_6$ octahedra in
Sr$_2$IrO$_4$. In order to solve the pseudospin Hamiltonian problem and evaluate the N\'eel temperature ($T_N$),
we employ the non-linear sigma model. We have found that, while for Sr$_2$IrO$_4$ our value of $T_N$ agrees with
the experimental data, for Ba$_2$IrO$_4$ it is overestimated by a factor two. We argue that this discrepancy is related to
limitations of the SE model: while for more localized $t_{2g}$ states in Sr$_2$IrO$_4$ it works reasonably well,
the higher-order terms in the perturbation theory expansion play a more important role in the more
``itinerant'' Ba$_2$IrO$_4$, giving rise to the new type of isotropic and anisotropic exchange interactions, which are
not captured by the SE model.
This conclusion is supported by unrestricted Hartree-Fock calculations for the same electron model, where in the case of Ba$_2$IrO$_4$,
already on the mean-field level,
we were able to reproduce the experimentally observed magnetic ground state, while for Sr$_2$IrO$_4$ the main results are essentially the same
as in the SE model.
\end{abstract}

\maketitle




\section{\label{sec:Intro} Introduction}
$5d$ transition-metal oxides have attracted a considerable attention as a new paradigm of
relativistic magnetic materials,
whose properties are largely influenced by the strong spin-orbit (SO) coupling, leading to the
experimental realization and a number of theoretical proposals for such fascinating phenomena as
SO interaction assisted Mott state in Sr$_2$IrO$_4$,\cite{Cao,BJKim1,BJKim}
spin-liquid state in Pr$_2$Ir$_2$O$_7$ (Ref.~\onlinecite{Nakatsuji})
and Na$_4$Ir$_3$O$_8$ (Refs.~\onlinecite{Okamoto,ChenBalents}),
possible existence of topological semimetallic states in
pyrochlore iridates,\cite{Wan}
and unusual magnetic ordering in the honeycomb compounds
Na$_2$IrO$_3$ and Li$_2$IrO$_3$,\cite{Liu,Choi,Biffin,Takayama} which may be relevant to the
Kitaev model of bond-dependent anisotropic magnetic coupling.\cite{Kitaev}

  In this respect, a lot of attention is being focused on the properties of tetravalent iridium oxides (or iridates),
originating from the $5/6$-filled Ir $t_{2g}$ band, located
near the Fermi level. The strong SO interaction splits the atomic $t_{2g}$ states into
the fully occupied fourfold degenerate $j=3/2$ states and twofold (Kramer's) degenerate $j=1/2$ states,
which accommodate one electron.
In this sense, there is a clear analogy with the spin-1/2 systems and the problem of interatomic
exchange interactions can be formulated in terms of some appropriately selected pseudospin states.
In solids, each group of states form the bands, which can, however, overlap with each other.
Moreover, since $j$ is the band quantum number in solids, there is always a finite hybridization
between these two groups of relativistic states.
The $j=1/2$ electrons experience the on-site Coulomb repulsion and can
polarize the occupied $j=3/2$ shell via the intraatomic exchange interactions.
Moreover, the precise division of the $t_{2g}$ states into the $j=3/2$ and $j=1/2$ ones
depends on the crystal-field splitting, which is typically smaller than the SO coupling.
These are the main ingredients, which predetermine the low-energy electronic properties of iridates.

  The layered perovskites, Sr$_2$IrO$_4$ and Ba$_2$IrO$_4$, are typically regarded as the key materials
for revealing the basic microscopic mechanisms, which can operate in iridates. They are also used as the
benchmark materials for testing the new theoretical models. In this respect, the first and one of the most
successful theoretical models for iridates was based on the
theory of superexchange (SE) interactions, which is valid in the limit of large on-site Coulomb repulsion and
treats the transfer integrals between the relativistic pseudospin states
in the second order of perturbation theory.\cite{ChenBalents,Khaliullin}
This model indeed reveals a rich and very interesting physics, including the bond dependence
of the anisotropic exchange couplings and emergence of large antisymmetric Dzyaloshinskii-Moriya (DM) interactions
when the inversion symmetry is broken by the anti-phase rotations of the IrO$_6$ octahedra in Sr$_2$IrO$_4$.

  At the same time, there was always a question about how far one can go in applying the SE model
for the real iridates. For the layered iridates, this point was risen in Ref.~\onlinecite{Arita},
where, using the dynamical mean-field theory (DMFT), the authors of
this work have argued that the behavior of both Sr$_2$IrO$_4$ and Ba$_2$IrO$_4$
retain many aspects of Slater insulators, whose insulating properties are closely related to the existence of the
long-range antiferromagnetic (AFM) order. The problem reemerged again recently after the experimental discovery of
the magnetic ground state structure of Ba$_2$IrO$_4$,\cite{Boseggia}
which cannot be described by the SE model, at least on the mean-field level.\cite{KatukuriPRX}
Thus, the questions is whether this problem should be resolved by considering the quantum fluctuation effects,
but within the SE model,\cite{KatukuriPRX} or
revising the SE model itself by including to it
some new higher-order terms in the perturbation theory expansion. The answer
to this question is not obvious,
because in the SE formulation, the effects of the SO coupling are included to the transfer integral. Therefore,
the second-order perturbation theory with respect to the transfer integrals automatically means that it treat
the SO coupling also only up to the second order. If the SO interaction is large (as in iridates), it can be
rather crude approximation, because it does not take into account several important effects, such as the in-plane
anisotropy in the uniaxial systems, which may be relevant to the experimentally
observed behavior of Sr$_2$IrO$_4$ and Ba$_2$IrO$_4$.
Another interesting point is the value of N\'eel temperature ($T_N$), which is remarkably close in both
considered systems (about $240$ K), and whether this fact can be rationalized on the basis
of SE theory.

  The main purpose of this work is to critically reexamine abilities of the SE theory for the layered iridates.
This is certainly not the first attempt to derive parameters of interatomic exchange interactions
using the theory of SE interactions
and the basic ideas of this method in the case of the strong SO coupling
are well understood today, at least for the models.\cite{Khaliullin,KatukuriPRX,Jin,BHKim,Igarashi,Perkins}
Nevertheless, besides the SO coupling,
the behavior of interatomic exchange interactions strongly depends on the
number of adjustable parameters, used in the model Hamiltonians, such as the on-site Coulomb and exchange interactions,
tetragonal crystal-field splitting, and the matrices of transfer integrals. Therefore, we believe
that, in the process of derivation of the pseudospin model, it is very important
to reduce the number of possible ambiguities by sticking as much as possible to the first-principles
electronic structure calculations.

  The rest of the article is organizes as follows. In Sec.~\ref{sec:Structure} we will briefly discuss the
main differences of the crystal and electronic structure of Ba$_2$IrO$_4$ and Sr$_2$IrO$_4$.
Then, in Sec.~\ref{sec:elmodel} we will explain our method of
the construction of the effective low-energy electron model
on the basis of first-principles electronic structure calculations with the SO coupling.
This model will be further used in Sec.~\ref{sec:psmodel} as the
staring point for the derivation of the SE Hamiltonian in the basis of
pseudospin states. In Sec.~\ref{sec:results} we will discuss results of our calculations of
the SE interactions and their implications to the magnetic properties of Ba$_2$IrO$_4$ and Sr$_2$IrO$_4$
using the non-linear sigma model.
Then, in Sec.~\ref{sec:beyondSE}, we will provide a detailed comparison with
the results of unrestricted Hartree-Fock (HF) calculations, which do not rely on the perturbation theory,
and argue that while for Sr$_2$IrO$_4$ the SE theory works reasonably well,
for Ba$_2$IrO$_4$  it misses several important interactions, which are nonetheless captured
by the HF calculations.
Finally, in Sec.~\ref{sec:conc}, we will give a brief summary of our work.
Details of derivation of the non-linear sigma model for the compass
Heisenberg model will be given in the Appendix.

\section{\label{sec:Structure} Main details of crystal and electronic structure}

\begin{figure}[tbp]
\begin{center}
\includegraphics[width=6cm]{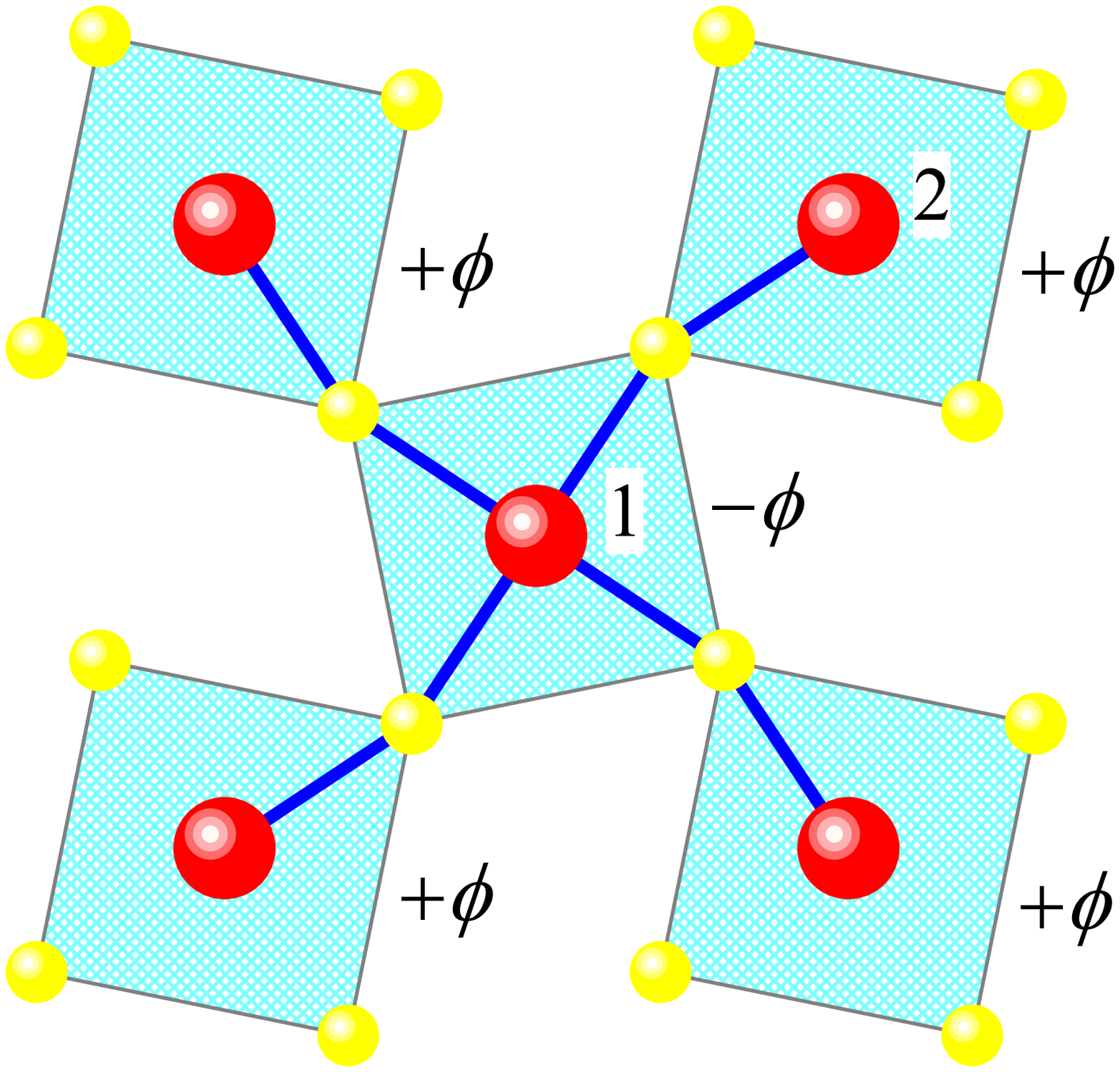} \qquad
\includegraphics[width=3cm]{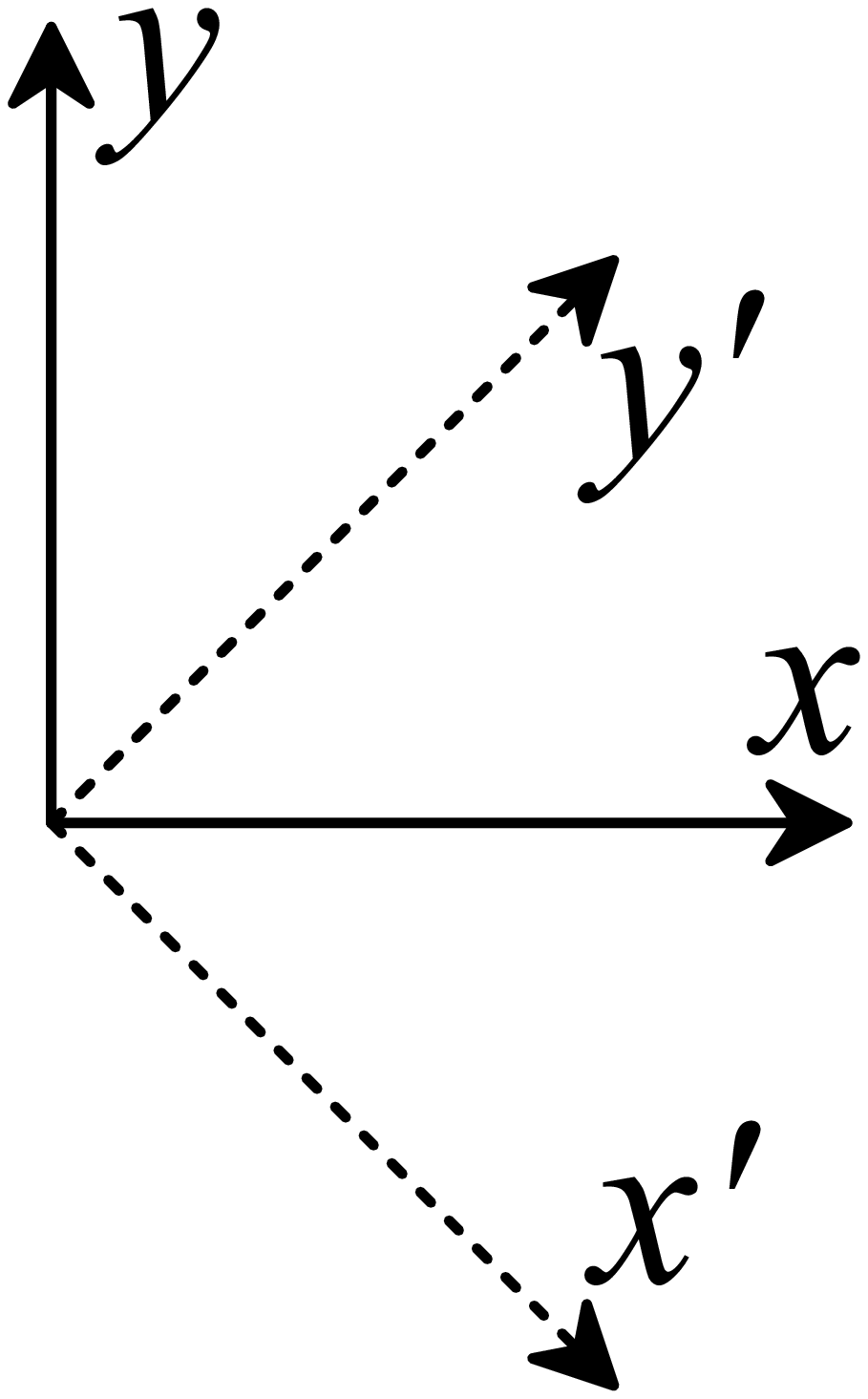}
\end{center}
\caption{(Color online) (Left) Rotations of IrO$_6$ octahedra in the $xy$ plane of Sr$_2$IrO$_4$.
The Ir atoms are indicated by the big (red) spheres and the oxygen
atoms are indicated by the small (yellow) spheres. The sites around which the octahedra are
rotated clockwise ($+$$\phi$) and counterclockwise ($-$$\phi$) are denoted as $1$ and $2$, respectively.
(Right)
The directions of axes in the $I4_1/acd$ $(x,y)$ and $4m/mmm$ $(x^{\prime},y^{\prime})$
coordinate frames.}
\label{fig.structure}
\end{figure}

  In this work we use the experimental structure
parameters, reported in Refs.~\onlinecite{Okabe} and \onlinecite{Crawford}  (at 13 K)
for Ba$_2$IrO$_4$ and Sr$_2$IrO$_4$, respectively. According to these data, Ba$_2$IrO$_4$
crystallizes in the undistorted tetragonal $4m/mmm$ structure with the Ir-O-Ir angles in the
$xy$ plane being equal to $180^\circ$. Sr$_2$IrO$_4$ exhibits the additional rotation of
IrO$_6$ octahedra (the space group $I4_1/acd$), which leads to the deformation of the
Ir-O-Ir bonds in the $xy$ plane (see Fig.~\ref{fig.structure}).
Depending on the Ir site, this rotation can be either clockwise ($+$$\phi$) or
counterclockwise ($-$$\phi$). The experimental value of the angle $\phi$ is $12^\circ$.\cite{Crawford}

\begin{figure}[tbp]
\begin{center}
\includegraphics[width=15cm]{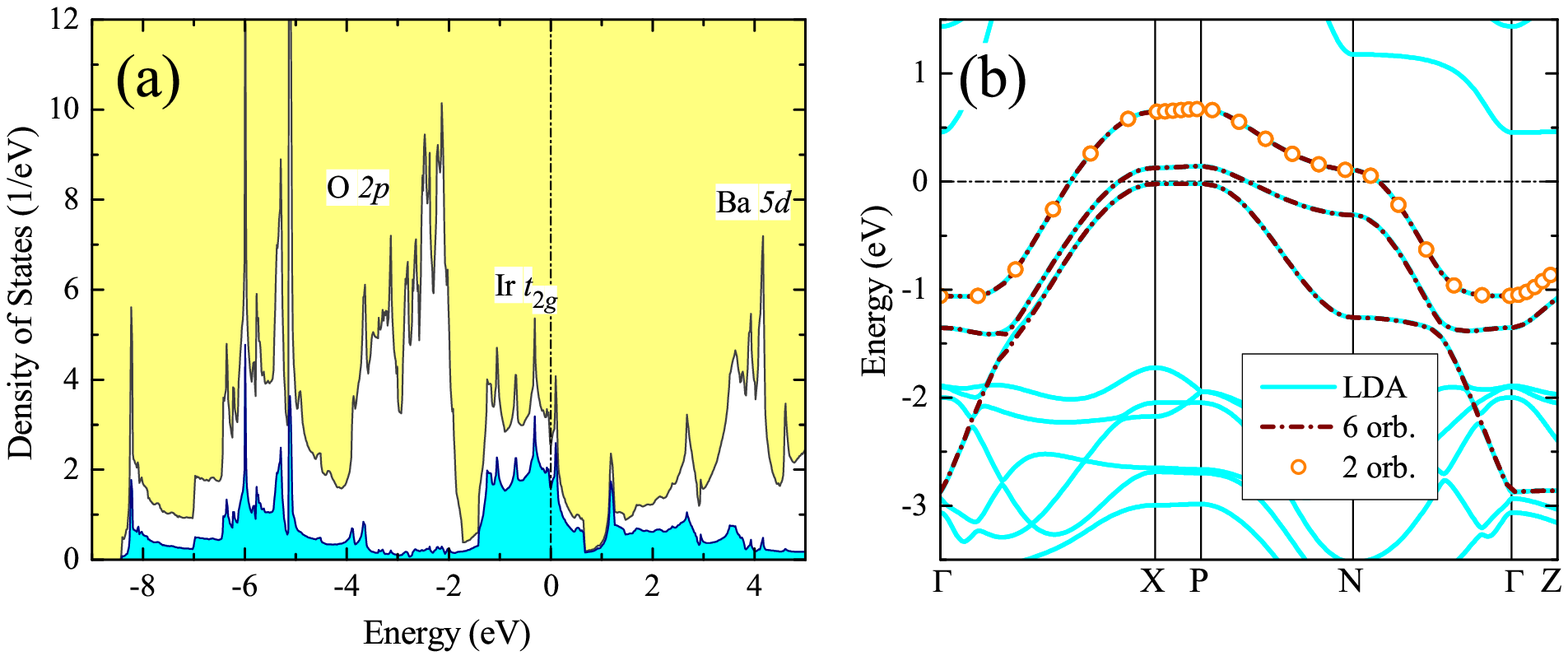}
\end{center}
\caption{(Color online) Electronic structure of Ba$_2$IrO$_4$ in LDA with
the SO coupling. (a) Total and partial densities of states. The shaded area
shows the contributions of the Ir $5d$ states. The positions of the main
bands are indicated by symbols. (b) Band dispersion near the Fermi level, as
obtained for the full LDA Hamiltonian in comparison with the six- and two-orbital
models. The high-symmetry points of the Brillouin zone are denoted as
$\Gamma = (0, 0, 0)$, $\mathrm{X} = (\protect\pi/a, \protect\pi/a, 0)$,
$\mathrm{N} = (\protect\pi/a, 0, \protect\pi/c)$,
$\mathrm{P} = (\protect\pi /a, \protect\pi/a, \protect\pi/c)$, and $\mathrm{Z} = (0, 0, 2\protect\pi/c)$.
The Fermi level is at zero energy (shown by dot-dashed line). }
\label{fig.BIODOS}
\end{figure}

\begin{figure}[tbp]
\begin{center}
\includegraphics[width=15cm]{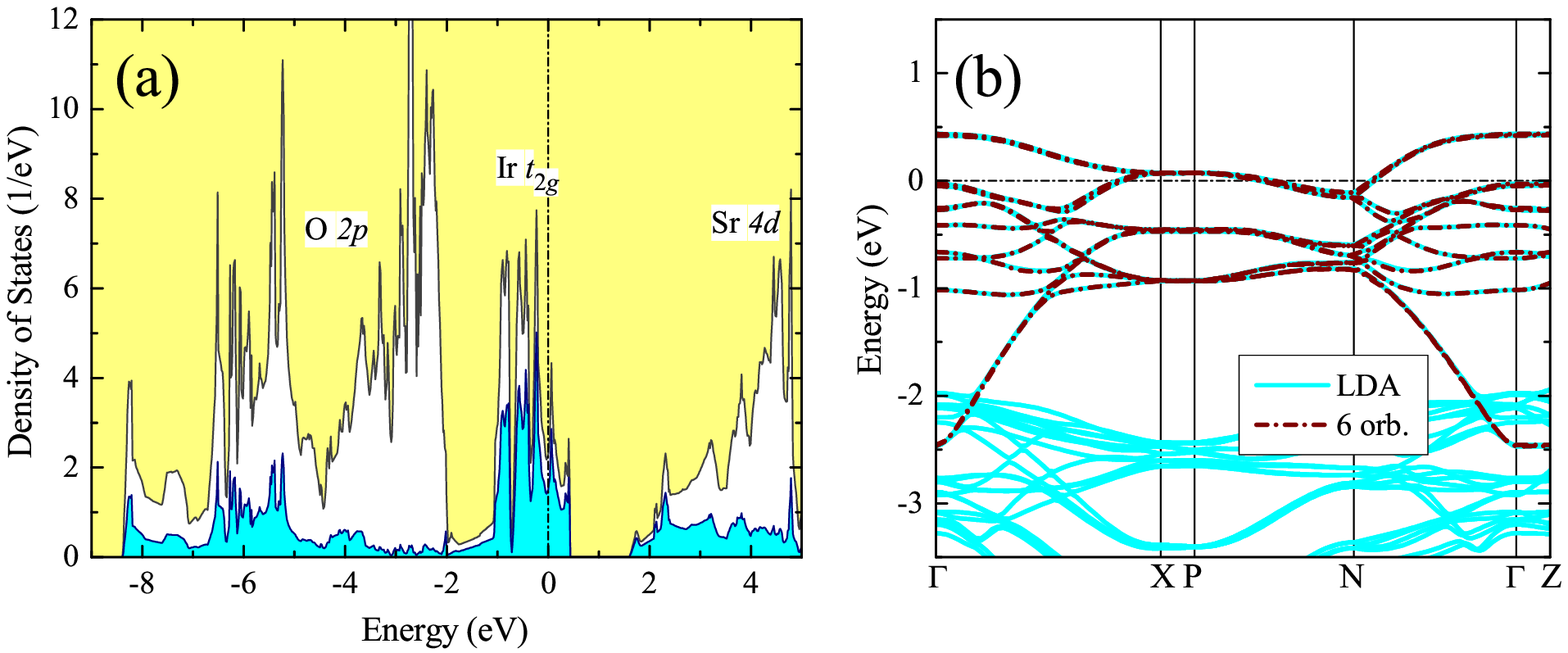}
\end{center}
\caption{(Color online) Electronic structure of Sr$_2$IrO$_4$ in LDA with
the SO coupling. (a) Total and partial densities of states. The shaded area
shows the contributions of the Ir $5d$ states. The positions of the main
bands are indicated by symbols. (b) Band dispersion near the Fermi level, as
obtained for the full LDA Hamiltonian in comparison the six-orbital models. The
Fermi level is at zero energy (shown by dot-dashed line). }
\label{fig.SIODOS}
\end{figure}

  The corresponding electronic structure in the local-density approximation (LDA) with the
SO coupling is displayed in Figs.~\ref{fig.BIODOS} and \ref{fig.SIODOS}
for Ba$_2$IrO$_4$ and Sr$_2$IrO$_4$, respectively.
In this work we will focus on the behavior of magnetically active Ir $t_{2g}$ bands, located near the Fermi level and
separated relatively well from the rest of the spectrum. There are two main differences between Ba$_2$IrO$_4$ and Sr$_2$IrO$_4$:
(i) The Ir $t_{2g}$ band is narrower in Sr$_2$IrO$_4$ (the bandwidth is about about $3$ and $3.5$ eV in Sr$_2$IrO$_4$ and
Ba$_2$IrO$_4$, respectively). This is generally consistent with the additional distortion in Sr$_2$IrO$_4$,
which leads to the deformation of the Ir-O-Ir bonds; (ii) The Ba $5d$ band in Ba$_2$IrO$_4$, which
strongly hybridizes and, therefore, has a large weight of
the Ir $5d$ states, is much closer to the Fermi energy than the Sr $4d$ band in Sr$_2$IrO$_4$.
This is mainly related to the larger Ba $5d$ bandwidth, due to the less distorted crystal structure
as well as the relativistic effects.\cite{Christensen}

  In Ba$_2$IrO$_4$, the relativistic $j=3/2$ and $j=1/2$ subbands are separated by the direct gap, which allows us to
construct both 6- and 2-orbitals models (for the entire $t_{2g}$ bands and $j=1/2$ subband, respectively).
In Sr$_2$IrO$_4$, due to the additional mixing between the $j=3/2$ and $j=1/2$ states, caused by the $I4_1/acd$ distortion,
such separation does not take place. Therefore, for Sr$_2$IrO$_4$, we will focus only on the 6-orbital model.

\section{\label{sec:elmodel} Construction of effective low-energy electron model}
In this section we will discuss the construction of the low-energy electron
model, starting from the LDA band structure with the SO interaction. For practical calculations, we
use the linear muffin-tin orbital (LMTO) method in the nearly orthogonal
representation.\cite{LMTO} The model itself has the following form:
\begin{equation}
\hat{\mathcal{H}}_{\mathrm{el}} = \sum_{ij} \sum_{\alpha \beta}
t_{ij}^{\alpha \beta}\hat{c}^\dagger_{i\alpha} \hat{c}^{\phantom{\dagger}}_{j\beta} +
\frac{1}{2} \sum_{i} \sum_{ \alpha \beta \gamma \delta }
U_{\alpha \beta \gamma \delta} \hat{c}^\dagger_{i\alpha} \hat{c}^\dagger_{i\gamma} \hat{c}^{\phantom{\dagger}}_{i\beta} \hat{c}^{\phantom{\dagger}}_{i\delta},  \label{eqn:Hel}
\end{equation}
where $\hat{c}^\dagger_{i\alpha}$ and $\hat{c}_{i\alpha}$ are, respectively, the creation
and annihilation operators of an electron on the Wannier orbitals $w_{i \alpha}$, centered at
the Ir site $i$ and specified by the index $\alpha = (m,s)$,
which numbers Kramer's doublets $m$$= 1$, $2$, or $3$ (an analog of
orbital indices without SO coupling) and the states $s$$=1$ or $2$ within
each such doublet (an analog of spin indices without SO coupling).

  First, we construct the Wannier functions for the magnetically active bands,
using the projector-operator technique.\cite{review2008,MarzariVanderbilt,PRB07}
We consider the 6-orbital model for the both
Ba$_2$IrO$_4$ and Sr$_2$IrO$_4$. Moreover, for Ba$_2$IrO$_4$ it is also
possible to construct the 2-orbital model, by considering only two highest
degenerate bands (see Fig.~\ref{fig.BIODOS}). The trial functions, which are
used for the projection, were obtained from the digonalization of
the site-diagonal density matrix, calculated for the magnetically
active bands in the basis of Ir $5d$ orbitals.\cite{review2008,PRB07}
Namely, after the diagonalization of the density matrix, we pick up either 6
or 2 eigenstates (depending on the dimensionality of the model) with the
largest eigenvalues and use them as the trial functions. Such
construction guarantees that the Wannier functions are well localized in
the real space: the main part of the density matrix with the largest
eigenvalues is described by the ``heads'' of the Wannier functions, residing
on the central site, and only small remaining part of this matrix is
described by the ``tails'' of the Wannier functions, coming from the
neighboring Ir sites. Thus, the main weight of the Wannier function is
concentrated in its ``head'' part, while the contribution
of ``tails'' is relatively small. Such procedure was extensively used in
nonrelativistic calculations without the SO coupling.\cite{review2008} The new
aspect of the relativistic formulation is that the eigenstates of the
density matrix become Kramers degenerate. Therefore, the trial functions
and the Wannier functions ($w_1$ and $w_2$) for each Kramer's doublet can be
chosen so to satisfy the conditions:
$| w_2 \rangle = \hat{T} | w_1 \rangle$ and $| w_1 \rangle = -$$\hat{T} | w_2 \rangle$, where
$\hat{T} = i \hat{\sigma}_y \hat{K}$ is the time-reversal operation, written in terms of the
spin Pauli matrix $\hat{\sigma}_y$ and the complex conjugation operator $\hat{K}$.

  Then, the one-electron part of the model Hamiltonian (\ref{eqn:Hel}) is
identified with the matrix elements of the LDA Hamiltonian in the Wannier basis: $t_{ij}^{\alpha \beta} = \langle w_{i
\alpha} | \hat{\mathcal{H}}_{LDA} | w_{j \beta} \rangle$. This procedure can
be also reformulated as the downfolding of the LDA Hamiltonian.\cite{review2008,PRB07}
Then, the site-diagonal matrix elements $t_{i = j}^{\alpha \beta}$
describe the splitting of the atomic levels by the crystal field and the SO
interaction, while the off-diagonal elements $t_{i \ne j}^{\alpha \beta}$ stand for interatomic transfer integrals
(or kinetic hoppings).

  The matrix of screened on-site interactions
$\hat{U} = [U_{\alpha \beta \gamma \delta}]$ has been calculated using simplified version of the
constrained random-phase approximation (RPA).\cite{review2008} The
RPA is used in the GW method in order to evaluate the momentum and frequency dependence
of the screened Coulomb interaction, which is then used in the calculations
of the self-energy.\cite{GW} The basic idea of constrained RPA is to switch off some
contributions to the RPA polarization function (and, therefore, to the
screening of $\hat{U}$) related to the transition between the magnetically
active bands (in our case, the Ir $5d$ bands).\cite{Ferdi04}
The RPA is inadequate for this channel of screening
(especially when it is evaluated starting from the LDA bandstructure) and
should be replaced by a more rigorous method in the process of solution of
the low-energy model (\ref{eqn:Hel}). The purpose of additional
simplifications is to replace the time-consuming RPA for the screening,
caused by the relaxation of the atomic wavefunction and other (non-$5d$) states, by much faster and more
suitable for these purposes constrained LDA technique. After that, we
consider (within RPA) the additional and most efficient channel of screening of the Coulomb
interactions in the Ir $5d$ bands by (the same) Ir $5d$ states, which contribute to
other parts of the electronic structure (mainly to the O $2p$ and either Ba $5d$ or Sr $4d$ bands in
Figs.~\ref{fig.BIODOS} and \ref{fig.SIODOS}) due to the hybridization.\cite{review2008} Such approximation
incorporates the main channels of screening and, thus, reproduces reasonably well
the values of static Coulomb interactions, obtained in full-scale
constrained RPA calculations. The obtained matrix elements $U_{\alpha \beta
\gamma \delta}$ have the following form:
\begin{equation}
U_{\alpha \beta \gamma \delta} = \int d \mathbf{r} \int d \mathbf{r}^{\prime}w_{\alpha}^\dagger (\mathbf{r}) w_{\beta} (\mathbf{r}) v_{\mathrm{scr}}(\mathbf{r},\mathbf{r}^{\prime}) w_{\gamma}^\dagger (\mathbf{r}^{\prime}) w_{\delta} (\mathbf{r}^{\prime}),  \label{eqn:scrU}
\end{equation}
where the screened interaction $v_{\mathrm{scr}}(\mathbf{r},\mathbf{r}^{\prime})$ in RPA is invariant under the time-reversal operation and does
not depend on the spin variables.

\section{\label{sec:psmodel} Pseudospin model}
In this section we will consider the mapping of the
electron model (\ref{eqn:Hel}) onto the magnetic model, formulated in terms of
\textit{pseudospin} variables
$\boldsymbol{ \cal S}_{i} = (\mathcal{S}_{i}^x, \mathcal{S}_{i}^y, \mathcal{S}_{i}^z)$:
\begin{equation}
\hat{\mathcal{H}}_{\mathcal{S}} = \sum_{i > j} \boldsymbol{ \cal S}_i
\tensor{\boldsymbol{J}}_{ij} \boldsymbol{ \cal S}_j +
\sum_{i} \boldsymbol{ \cal S}_i \tensor{\boldsymbol{g}}_i \boldsymbol{ H},
\label{eqn:HS}
\end{equation}
where $\tensor{\boldsymbol{J}}_{ij}$ and $\tensor{\boldsymbol{g}}_i$ are the
$3$$\times$$3$ tensors, describing interactions in the system of pseudospins
and with the external magnetic field $\boldsymbol{H}$, respectively. The
pseudospin operators are represented by the Pauli matrices:
$ \mathcal{S}_{i}^x = \frac{1}{2} \left(
\begin{array}{cc}
0 & 1 \\
1 & 0
\end{array}
\right)$, $ \mathcal{S}_{i}^y = \frac{1}{2} \left(
\begin{array}{cc}
0 & -i \\
i & 0
\end{array}
\right)$, and $ \mathcal{S}_{i}^z = \frac{1}{2} \left(
\begin{array}{cc}
1 & 0 \\
0 & -1
\end{array}
\right)$.

  For each bond, $\tensor{\boldsymbol{J}}_{ij}$ can be presented as the sum of
its symmetric ($S$) and antisymmetric ($A$) components:
$\tensor{\boldsymbol{J}}_{ij} = \tensor{\boldsymbol{J}}^{(S)}_{ij} + \tensor{\boldsymbol{J}}^{(A)}_{ij}$,
where $\tensor{\boldsymbol{J}}^{(S)}_{ij} = \frac{1}{2}(\tensor{\boldsymbol{J}}_{ij} +
\tensor{\boldsymbol{J}}^T_{ij})$ and $\tensor{\boldsymbol{J}}^{(A)}_{ij} =
\frac{1}{2}(\tensor{\boldsymbol{J}}_{ij} - \tensor{\boldsymbol{J}}^T_{ij})$.
The part $\tensor{\boldsymbol{J}}^{(S)}_{ij}$ incorporates all types of symmetric
exchange interactions and its trace is the isotropic exchange interaction in
the bond $i$-$j$: $J_{ij} = \mathrm{Tr} \tensor{\boldsymbol{J}}^{(S)}_{ij}$,
while $\tensor{\boldsymbol{J}}^{(A)}_{ij}$ describes anisotropic
DM interactions. $\tensor{\boldsymbol{J}}^{(A)}_{ij}$
has only three independent elements, which can be viewed as the components
of some axial vectors (the so-called DM vector) $\boldsymbol{d}_{ij} =
(d_{ij}^x,d_{ij}^y,d_{ij}^z)$:
\begin{equation*}
\tensor{\boldsymbol{J}}^{(A)}_{ij} = \left(
\begin{array}{ccc}
0 & d^z_{ij} & -d^y_{ij} \\
-d^z_{ij} & 0 & d^x_{ij} \\
d^y_{ij} & -d^x_{ij} & 0
\end{array}
\right),
\end{equation*}
yielding the well known identity:
$\boldsymbol{ \cal S}_i \tensor{\boldsymbol{J}}^{(A)}_{ij} \boldsymbol{ \cal S}_j =
\boldsymbol{d}_{ij} [\boldsymbol{ \cal S}_i \times \boldsymbol{\cal S}_j]$.

\subsection{\label{sec:SEbasis} Calculation of superexchange interactions}
In order to calculate the SE interactions, we
adapt a standard procedure for the systems,
whose degeneracy in the atomic limit is lifted by the crystal field and SO interaction. Namely, we
assume that, in the atomic limit, the single hole resides on the highest Kramer's doublet,
obtained from the diagonalization of the site-diagonal part $\hat{t} = [t^{\alpha \beta}_{i=j} ]$
of the one-electron Hamiltonian. The states $| \varphi_1 \rangle$ and $| \varphi_2 \rangle$ of
this Kramer's doublet are used for the construction of eigenstates
$|$$\pm$$x$$\rangle$, $|$$\pm$$y$$\rangle$, and $|$$\pm$$z$$\rangle$ of the
pseudospin operators $\mathcal{S}^x$, $\mathcal{S}^y$, and $\mathcal{S}^z$, respectively.
For convenience, we choose the
phases of these states so that $| \varphi_2 \rangle = \hat{T} | \varphi_1
\rangle$ and $| \varphi_1 \rangle = -$$\hat{T} | \varphi_2 \rangle$.

  Let us first explain the construction for $|$$\pm$$z$$\rangle$. For these purposes, one can choose
any pair of states, which is obtained by the unitary transformation of
$| \varphi_1 \rangle$ and $| \varphi_2 \rangle$. Moreover, since the states are
degenerate, the transformation will not change the total energy, and the
model (\ref{eqn:HS}) will not contain the single-ion anisotropy term. Then,
we employ the fact that, despite some complications caused by the strong SO
coupling, the magnetic moment will always have a finite spin component,
and define the pseudospin states $|$$+$$z$$\rangle$ and $|$$-$$z$$\rangle$
as those corresponding to, respectively,
the maximal and minimal projections of spin onto the $z$ axis. The problem
is equivalent to the diagonalization of the $2$$\times$$2$ spin Pauli matrix
$\hat{\sigma}_z$ in the basis of $| \varphi_1 \rangle$ and
$| \varphi_2 \rangle$.

  Then, one can readily define two other groups of states as
$|$$\pm$$x$$\rangle = \frac{1}{\sqrt{2}} |$$+$$z$$\rangle \pm \frac{1}{\sqrt{2}} |$$-$$z$$\rangle$ and
$|$$\pm$$y$$\rangle = \pm \frac{1-i}{2} |$$+$$z$$\rangle + \frac{1+i}{2} |$$-$$z$$\rangle$.
In this construction, the phases of $|$$\pm$$z$$\rangle$ were chosen to satisfy the condition:
$\hat{T} |$$-$$z$$\rangle = |$$+$$z$$\rangle$ and $\hat{T} |$$+$$z$$\rangle = -|$$-$$z$$\rangle$.
It allows us to define unambiguously all phases of $|$$\pm$$z$$\rangle$ but $\zeta$, which transforms
$|$$\pm$$z$$\rangle$ to $e^{\mp i \zeta}|$$\pm$$z$$\rangle$.
The latter phase is defined so to satisfy the condition $\langle$$+$$x| \hat{\sigma}_y |$$+$$x$$\rangle = 0$.

  In order to find $\tensor{\boldsymbol{J}}_{ij}$, we evaluate the energy gain
$\mathcal{T}_{ij}(a,b)$, caused by the virtual excitations of the hole from
the $a$-th orbital of the site $i$ to the $b$-th orbital of the site $j$ and
vice versa, in the second order of perturbation theory with respect to the
transfer integrals $t^{\alpha \beta}_{i \ne j}$. The denominators in
the SE theory are given by the energies of charge excitations $d_i^5 d_j^5
\rightarrow d_i^4 d_j^6$, which are the
energies of the two-hole states. In the process of virtual excitations, the
Pauli exclusion principle was guaranteed by the projection operators, which
permit the hoppings only between occupied and unoccupied orbitals. Moreover, for
the excited two-hole states, the problem was solved in the true many-body
fashion, by finding the eigenstates and the eigenenergies from the
diagonalization of the Coulomb interaction matrix $\hat{U}$ in the basis of
$\frac{6 \times 5}{2} = 15$ Slater determinants, constructed from 6 atomic orbitals. This
is a step beyond the mean-field approximation, which additionally stabilizes
the AFM interactions.\cite{review2008} Then, we consider
all combinations of $a$ and $b$ $=$ $\pm x$, $\pm y$, or $\pm z$, and map
the obtained energy gains onto the pseudospin model (\ref{eqn:HS}) for $\boldsymbol{ H} = 0$.
This procedure was discussed in details in Ref.~\onlinecite{NJP09}.

\subsection{\label{sec:gtensor} Calculation of $g$-tensor}
The $g$-tensor describes the interaction of the pseudopsin with the external
magnetic field [see Eq.~(\ref{eqn:HS})]. It can be evaluated using
Eq.~(31.34) of Ref.~\onlinecite{AshcroftMermin}, from which one can
find all 9 elements of the tensor $\tensor{\boldsymbol{g}}$ at each site of the lattice:
$\langle$$+$$z$$|(L_x+\sigma_x)|$$+$$z$$\rangle = g^{xz}$, $\langle$$+$$z$$|(L_y+\sigma_y)|$$+$$z$$\rangle = g^{yz}$, $\langle$$+$$z$$|(L_z+\sigma_z)|$$+$$z$$\rangle = g^{zz}$,
$\langle$$-$$z$$|(L_x+\sigma_x)|$$+$$z$$\rangle = g^{xx} + ig^{xy}$, $\langle$$-$$z$$|(L_y+\sigma_y)|$$+$$z$$\rangle = g^{yx} + ig^{yy}$,
and $\langle$$-$$z$$|(L_z+\sigma_z)|$$+$$z$$\rangle = g^{zx} + ig^{zy}$,
where $L_x$, $L_y$, and
$L_z$ are the matrices of angular momentum operators in the Wannier
basis. It is easy to separate
the spin $\tensor{\boldsymbol{g}}_S$ and orbital $\tensor{\boldsymbol{g}}_L$ contributions to the $g$-tensor, by considering
the matrix elements of only $\boldsymbol{\sigma}$ and $\boldsymbol{L}$, respectively.

\section{\label{sec:results} Results and discussions}

\subsection{\label{sec:2BIO} Two-orbital model for Ba$_2$IrO$_4$}
The two-orbital model is the simplest model,
which can be considered. In Ba$_2$IrO$_4$, the ``$j=1/2$'' bands are separated
from the rest of the spectrum
(see Fig.~\ref{fig.BIODOS})
and the construction is rather straightforward.

The form of transfer integrals in this case is very simple.
Since $\hat{t}$ is hermitian, each $2$$\times$$2$ matrix
$\hat{t}_{ij} = [ \hat{t}_{ij}^{\alpha \beta} ]$ satisfies the property:
$\hat{t}_{ji} = \hat{t}_{ij}^\dagger$. Then, since all Ir sites are located in the
inversion centers and connected by the translations, it holds $\hat{t}_{ji} = \hat{t}_{ij}$ and,
therefore, $\hat{t}_{ij} = \hat{t}_{ij}^\dagger$.
Finally, since $\hat{\mathcal{H}}_{LDA}$ is invariant under the
time-reversal operation, we will have two more identities:
$(t_{ij}^{11})^* = t_{ij}^{22}$ and $(t_{ij}^{12})^* = - t_{ij}^{21}$, which can be obtained
from
$(t_{ij}^{\alpha \beta})^* = \langle \hat{T} w_{i \alpha} | \hat{T} \hat{\mathcal{H}}_{LDA} | w_{j \beta} \rangle$.
Thus, in the
two-orbital model, each $\hat{t}_{ij}$
is proportional to the unity matrix
$\hat{t}_{ij} = t_{ij} \hat{1}$  in the subspace spanned by the indices
$\alpha (\beta) =$ $1$ and $2$, where $t_{ij}$ is a real constant.

\begin{figure}[tbp]
\begin{center}
\includegraphics[width=5cm]{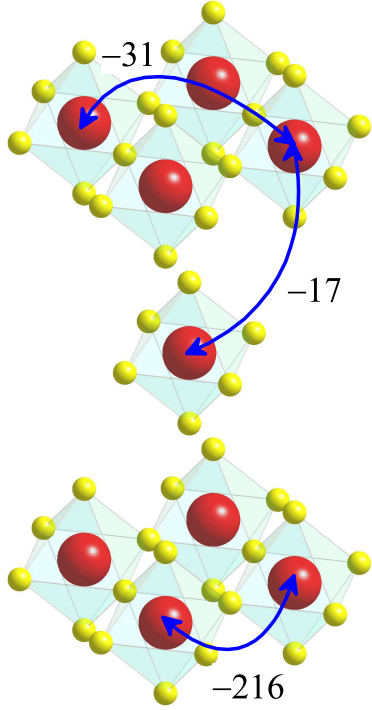}
\end{center}
\caption{(Color online) Crystal structure and transfer integrals (in meV)
associated with different Ir-Ir bonds in the two-orbital model for Ba$_2$IrO$_4$.
The Ir atoms are indicated by the big (red) spheres and the oxygen
atoms are indicated by the small (yellow) spheres. }
\label{fig.2oBIO}
\end{figure}

  The behavior of $t_{ij}$ is explained in Fig.~\ref{fig.2oBIO}.
As expected, the strongest hopping occurs between nearest neighbors in the $xy$ plane.
There are also finite hoppings between next-nearest
neighbors in and between the planes.

  Since $\hat{t}_{ij} = t_{ij} \hat{1}$, all SE interactions in the
two-orbital model are isotropic. They can be easily evaluated
using the formula $J_{ij} = 4t_{ij}^2/\mathcal{U}$,\cite{PWA}
where $\mathcal{U} = 1.52$ eV is the effective on-site Coulomb repulsion,
obtained in the constrained RPA for the two-orbital model. Then, using the
values of transfer integrals, which are displayed in Fig.~\ref{fig.2oBIO}, we will obtain
$J_{ij} =$ $122.8$, $2.5$, and $0.8$ meV for the nearest-neighbor (NN), next-NN,
and interplane interactions, respectively. Since $J_{ij} > 0$,
all interactions are antiferromagnetic.

\subsection{\label{sec:6BIO} Six-orbital model for Ba$_2$IrO$_4$}
The atomic $t_{2g}$ states are split into three doubly degenerate groups
of levels, which in Ba$_2$IrO$_4$ are located at $-$$209$, $-$$149$, and $358
$ meV, relative to their center of gravity. Two lowest doublets correspond to
$j=3/2$, and the highest one -- to $j=1/2$. Thus, the splitting between the
$j=1/2$ and $j=3/2$ states, which measures the strength of the SO coupling is
very large. This justifies the use of the regular (nondegenerate)
theory for the SE interactions.

  For the tetragonal compounds, the eigenstates $|$$+$$z$$\rangle$ (and $|$$-$$z$$\rangle = -$$\hat{T}|$$+$$z$$\rangle$),
corresponding to the highest Kramer's doublet,
can be decomposed in the basis
of $xy$, $yz$, $zx$, and $x^2$$-$$y^2$ Wannier orbitals with
both projection of spins:
\begin{eqnarray}
| +z \rangle & = & c^{\uparrow}_{xy} | w_{xy, \uparrow} \rangle +
c^{\uparrow}_{yz} | w_{yz, \uparrow} \rangle + c^{\uparrow}_{zx} | w_{zx,
\uparrow} \rangle + c^{\uparrow}_{x^2-y^2} | w_{x^2-y^2, \uparrow} \rangle +
\notag \\
& & c^{\downarrow}_{xy} | w_{xy, \downarrow} \rangle + c^{\downarrow}_{yz} |
w_{yz, \downarrow} \rangle + c^{\downarrow}_{zx} | w_{zx, \downarrow}
\rangle + c^{\downarrow}_{x^2-y^2} | w_{x^2-y^2, \downarrow} \rangle.
\label{eqn:z+}
\end{eqnarray}
Due to the symmetry constraint, the $3z^2$$-$$r^2$ orbitals do not
contribute to $|$$+$$z$$\rangle$. The coefficients in this expansion
depend on the relative strength of the crystal-field splitting and the SO
interaction. They cannot be determined solely from the symmetry principles.
For Ba$_2$IrO$_4$, we obtain the following (nonvanishing) coefficients in
the original $I4/mmm$ coordinate frame: $c^{\downarrow}_{x^{\prime}y^{
\prime}} = -$$i 0.522$, $c^{\uparrow}_{z^{\prime}x^{\prime}} = -$$i
c^{\uparrow}_{y^{\prime}z^{\prime}} = 0.603$, and
$c^{\downarrow}_{x^{\prime 2}-y^{\prime 2}} = -$$0.004$,
which correspond to $c^{\downarrow}_{xy} = 0.004$,
$c^{\uparrow}_{zx} = -$$i c^{\uparrow}_{yz} = 0.426 + i0.426$,
and $c^{\downarrow}_{x^2-y^2} = -$$i 0.522$ in the $I4_1/acd$ frame.

\begin{figure}[tbp]
\begin{center}
\includegraphics[width=10cm]{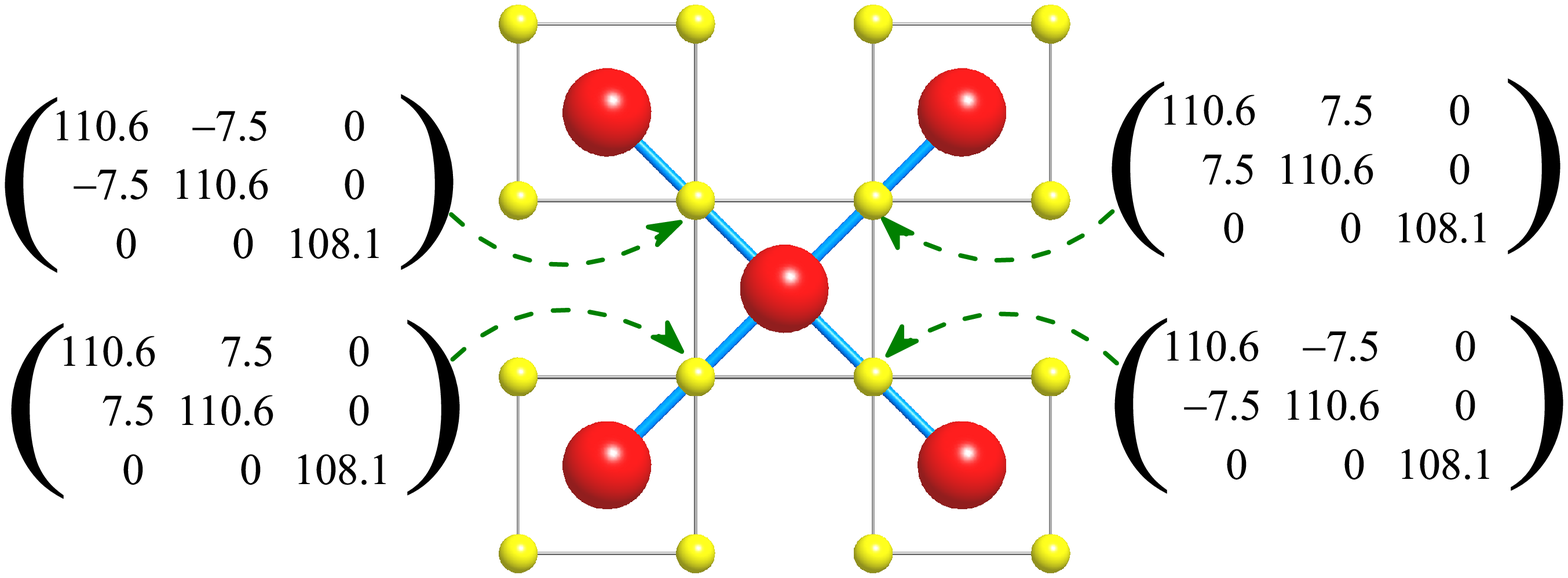}
\includegraphics[width=10cm]{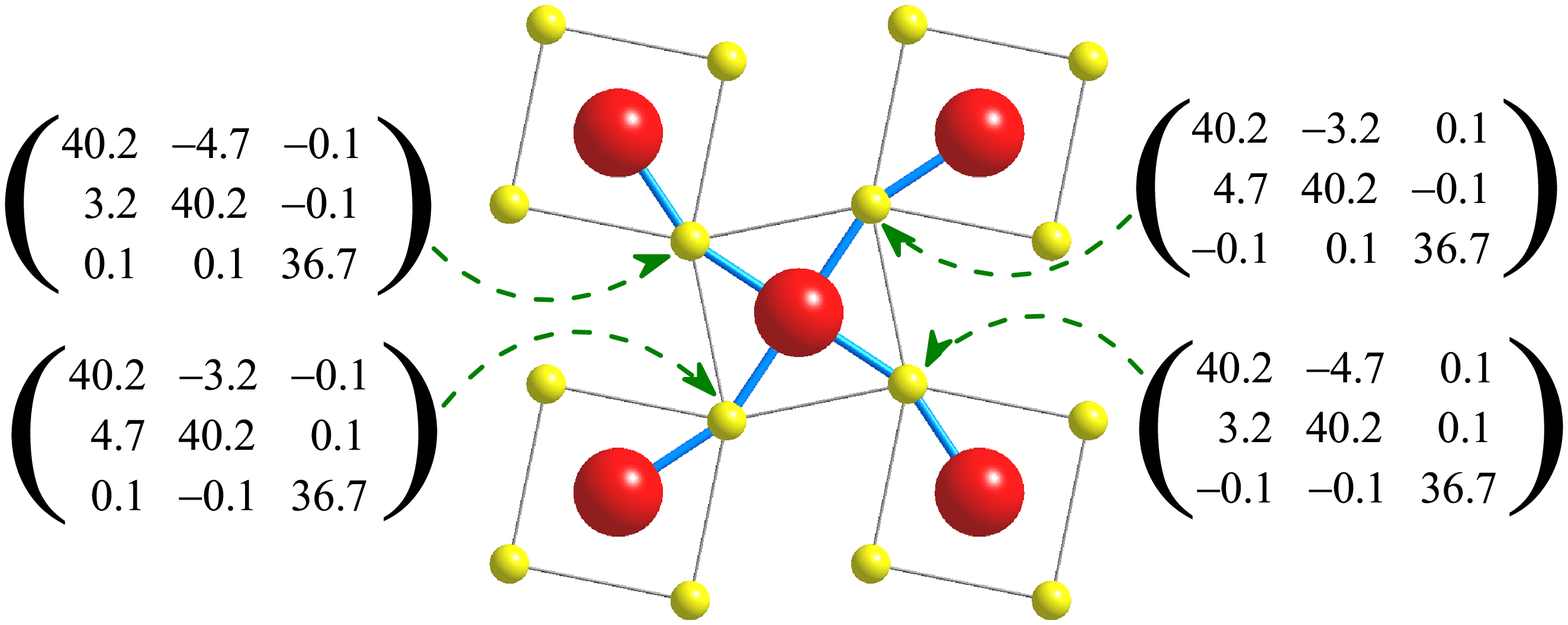}
\end{center}
\caption{(Color online) Tensors of superexchange interactions
$\tensor{\boldsymbol{J}}_{ij}$ (in meV), associated with different Ir-Ir
bonds in the $xy$ plane of Ba$_2$IrO$_4$ (top) and Sr$_2$IrO$_4$ (bottom).
The Ir atoms are indicated by the big (red) spheres and the oxygen atoms are
indicated by the small (yellow) spheres.
For the sake of convenience, the parameters for
both structures are shown in the $I4_1/acd$ coordinate frame. }
\label{fig.6oPlane}
\end{figure}

  The strongest transfer integrals, operating between the nearest neighbors in
the $xy$ plane, have the following form (in meV):
\begin{equation}
\hat{t}_{\langle ij \rangle || x^{\prime},y^{\prime}} = \left(
\begin{array}{cccccc}
-283 & 0 & 0 & \pm 60 & 0 & -i76 \\
0 & -283 & \mp 60 & 0 & -i76 & 0 \\
0 & \mp 60 & -165 & 0 & \pm i92 & 0 \\
\pm 60 & 0 & 0 & -165 & 0 & \mp i92 \\
0 & i76 & \mp i92 & 0 & -226 & 0 \\
i76 & 0 & 0 & \pm i92 & 0 & -226
\end{array}
\right),
\label{eqn:tij_BIO}
\end{equation}
where the upper (lower) sign stands for the bonds parallel to the
$x^{\prime}$ ($y^{\prime}$) axis in the $I4/mmm$ coordinate frame (see Fig.~\ref{fig.6oPlane}).
Here, the matrix is given in the local representation, which diagonalizes
the site-diagonal part of the one-electron Hamiltonian $[\hat{t}_{i=j}]$,
as described in Sec.~\ref{sec:SEbasis}.
Moreover, we adapt the following order of the Wannier orbitals:
$(m,s)$$=$ $(1,1)$, $(1,2)$, $(2,1)$, $(2,2)$, $(3,1)$, and $(3,2)$, where $m$
numbers the Kramer's doublet in the increasing order of their energies and $s
$ number the states within each doublet. Similar to the 2-orbital model,
the matrix elements of $\hat{t}_{ij}$ with same $m$ do not depend on the $s$-indices
and each such sub-block is proportional to the $2$$\times$$2$ unity matrix.
However, there is a finite
coupling between states with different $m$'s. This coupling gives rises to
the anisotropy of $\tensor{\boldsymbol{J}}_{ij}$. Moreover, since the signs
of some of these matrix elements alternate between the bonds parallel to the
$x^{\prime}$ and $y^{\prime}$ axes, the anisotropic part of $\tensor{\boldsymbol{J}}_{ij}$
will also alternate in the $x^{\prime}y^{\prime}$ plane.
Another important factor, which is responsible for anisotropic properties of
$\tensor{\boldsymbol{J}}_{ij}$ is the intraatomic exchange interaction $\mathcal{J}$.\cite{Khaliullin}
It will be discussed below.
Other parameters of the model Hamiltonian can be found elsewhere.\cite{SM}

\begin{figure}[tbp]
\begin{center}
\includegraphics[height=6cm]{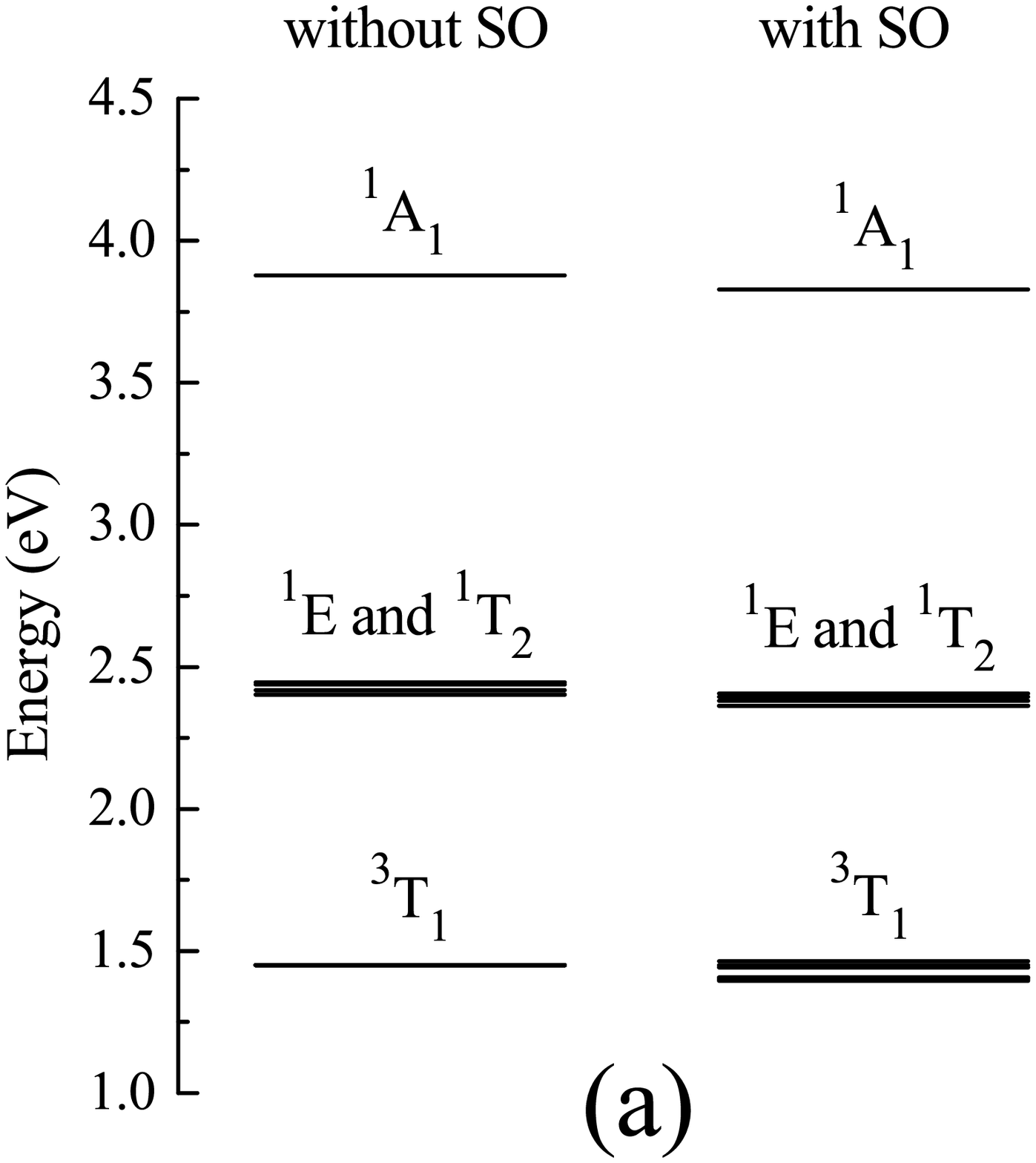}
\includegraphics[height=6cm]{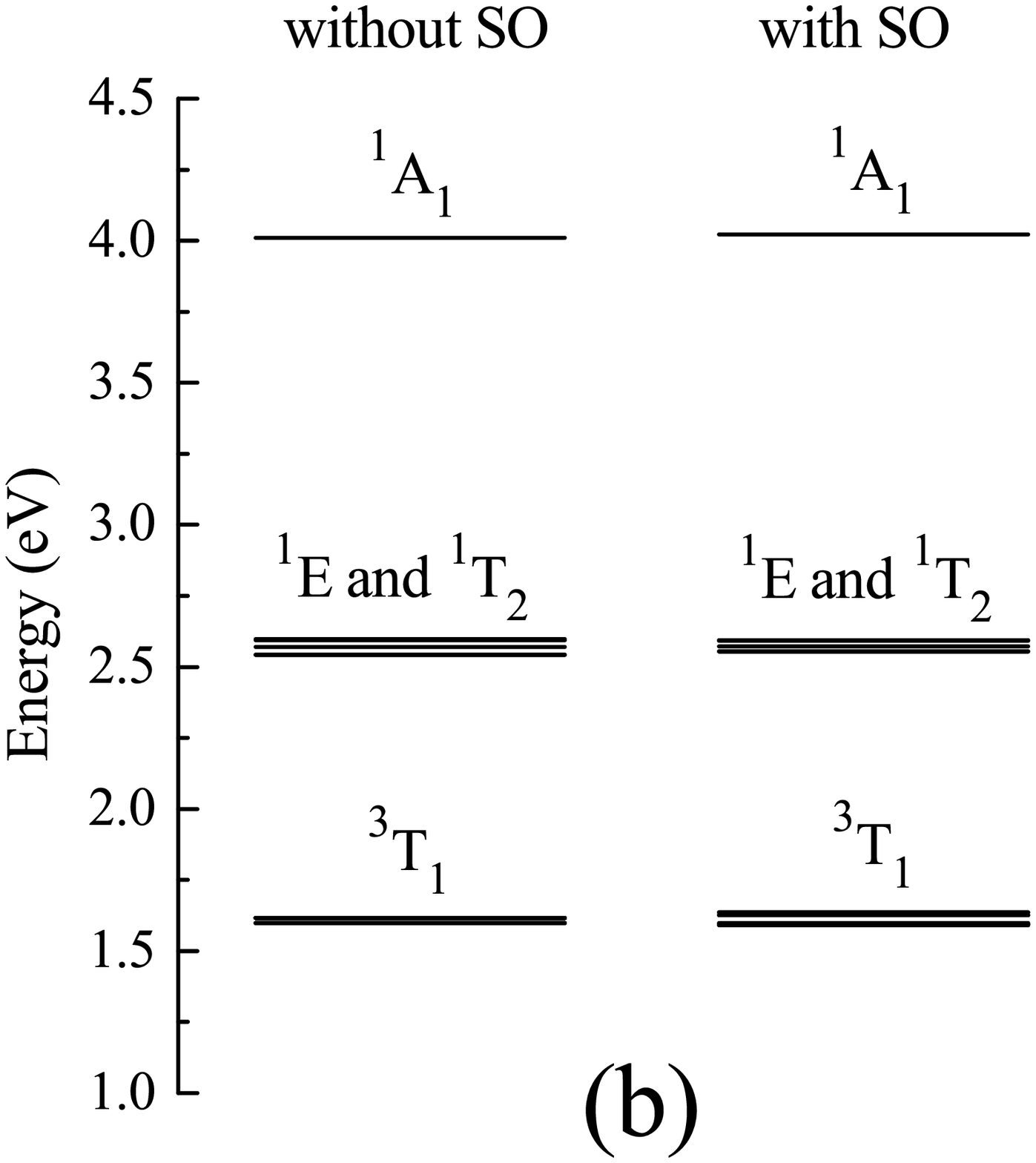}
\end{center}
\caption{The energies of two-hole states for Ba$_2$IrO$_4$
(a) and Sr$_2$IrO$_4$ (b), obtained using
parameters of screened Coulomb interactions $U_{\protect\alpha \protect\beta \protect\gamma \protect\delta}$
for the six-orbital model with and without the spin-orbit (SO) coupling. }
\label{fig.multiplet}
\end{figure}

  The form of the screened on-site interactions $U_{\alpha \beta \gamma \delta}$
in the basis of relativistic Wannier orbitals is rather complex.
Nevertheless, the main details of these interactions can be understood by
considering the energies of two-hole excitations, which contribute to the SE
processes (see Fig.~\ref{fig.multiplet}).
These energies were calculated using the matrices of screened Coulomb
interactions $[ U_{\alpha \beta \gamma \delta} ]$,
for which $v_{\mathrm{scr}}(\mathbf{r},\mathbf{r}^{\prime})$ was obtained for two types of the
electronic structures: with and without the SO coupling [see Eq.~(\ref{eqn:scrU})].
In the case of perfect cubic environment and without the SO
coupling, the two-hole states are split into three groups: $^3\mathrm{T}_1$,
degenerate $^1\mathrm{T}_2$ and $^1\mathrm{E}$, and $^1\mathrm{A}_1$ with
the energies $(\mathcal{U}$$-$$3\mathcal{J})$,
$(\mathcal{U}$$-$$\mathcal{J})$, and $(\mathcal{U}$$+$$2\mathcal{J})$, respectively,\cite{Oles2005} in
terms of the intraorbital Coulomb interaction $\mathcal{U}$ and the exchange
interaction $\mathcal{J}$.\cite{Kanamori} The tetragonal environment of the
Ir$^{4+}$ ions, realized in Ba$_2$IrO$_4$, slightly lifts the degeneracy of
the $^1\mathrm{T}_2$ and $^1\mathrm{E}$ states. The SO interaction further
lifts the degeneracy of the $^3\mathrm{T}_1$ states. However, in all other
respects the positions of the main energy levels in very similar with and
without the SO interactions. The (averaged) parameters $\mathcal{U}$ and $\mathcal{J}$
can be evaluated from the centers of gravity of the three
groups of levels. This yields:
$\mathcal{U} = 2.86$ ($2.91$) eV and $\mathcal{J} = 0.48$ ($0.49$) eV with (without) SO
interaction. Thus, $\mathcal{U}$ is generally larger in the 6-orbital
model, in comparison with the 2-orbital one, due to the additionally
screened by the $j=3/2$ electrons, which is included in the 2-orbital
model, but not in the 6-orbital one.

  The parameters of NN SE interactions in the $xy$ plane are explained in Fig.~\ref{fig.6oPlane}.
Since $J^{xx}_{ij} = J^{yy}_{ij} > J^{zz}_{ij}$, these
parameters favor the inplane configuration of the pseudospins, in agreement with
the experiment.\cite{Boseggia} Moreover, the phase of the off-diagonal
element $J^{xy}_{ij}$ (in the $I4_1/acd$ coordinate frame) is bond-dependent,
giving rise to the quantum compass interaction term. In the more
conventional $I4/mmm$ coordinate frame, the tensor $\tensor{\boldsymbol{J}}_{ij}$ is diagonal with the parameters given by
$J^{x^{\prime}x^{\prime}}_{ij} = J^{xx}_{ij} \pm |J^{xy}_{ij}|$ and
$J^{y^{\prime}y^{\prime}}_{ij} = J^{xx}_{ij} \mp |J^{xy}_{ij}|$, where the upper (lower) sign
stands for the bonds parallel to the $x^{\prime}$ ($y^{\prime}$) axis. The
isotropic part
$J_{ij} = \frac{1}{3}(J^{xx}_{ij}$$+$$J^{yy}_{ij}$$+$$J^{zz}_{ij}) = 109.8$ meV is
close to the value $J_{ij} = 123$ meV, obtained
in the 2-orbital model. This is mainly because of the combination of two
effects: On the one hand, $\mathcal{U}$ is larger in the
6-orbital model, which should lead to the smaller $J_{ij}$.
This decrease of $J_{ij}$ is partly compensated by somewhat stronger transfer
integrals, operating between orbitals belonging to the highest Kramer's
doublet ($-$$226$ meV instead of $-$$216$ meV in the two-orbital model).

   As was already mensioned before,
there are two important factors, which leads to the anisotropy of $\tensor{\boldsymbol{J}}_{ij}$:
(i) finite transfer integrals, connecting the states with $j=3/2$ and $j=1/2$ [see Eq.~\ref{eqn:tij_BIO}] and
(ii) finite intraatomic exchange coupling $\mathcal{J}$,\cite{Khaliullin}
which lifts the main degeneracy of the virtual two-hole states (see Fig.~\ref{fig.multiplet}).
For instance, using the same transfer integrals, but simplified matrix of the screened
on-site Coulomb interactions, which was reconstructed from the parameters of averaged
$\mathcal{U} = 2.86$ eV and $\mathcal{J} = 0$, we have obtained totally isotropic tensor
$\tensor{\boldsymbol{J}}_{ij} \equiv J_{ij} \tensor{\boldsymbol{1}}$, where
$\tensor{\boldsymbol{1}}$ is the $3$$\times$$3$ unity tensor and $J_{ij} = 71$ meV.

  The direction of the uniaxial anisotropy is also controlled by the tetragonal crystal-field splitting
$\Delta_{t_{2g}}$
between $x^{\prime}y^{\prime}$ and doubly degenerate
$y^{\prime}z^{\prime}$ and $z^{\prime}x^{\prime}$ orbitals without the SO coupling.
If the $x^{\prime}y^{\prime}$ orbital was located higher in energy,
we would deal with the out-of-plane configuration of pseudospins: $J^{xx}_{ij} = J^{yy}_{ij} < J^{zz}_{ij}$.
Such situation is indeed realized for $\Delta_{t_{2g}} = 107$ meV, associated entirely
with the change of the hybridization due to the compression of the IrO$_6$ octahedra in the $x^{\prime}y^{\prime}$ plane.
However, the additional nonspherical Madelung interaction (see Ref.~\onlinecite{review2008})
yields $\Delta_{t_{2g}} = -$$69$ meV, thus
changing the order of the $t_{2g}$ orbitals
and enforcing the in-plane configuration of the pseudospins
($J^{xx}_{ij} = J^{yy}_{ij} > J^{zz}_{ij}$), in agreement with the
experimental data.\cite{Boseggia}
In mathematical terms, it leads to the inequality
$|c^{\downarrow}_{x^{\prime}y^{\prime}}| < |c^{\uparrow}_{y^{\prime}z^{\prime}}|$$=$$|c^{\uparrow}_{z^{\prime}x^{\prime}}|$
for the coefficients in Eq.~(\ref{eqn:z+}).
It is interesting that for the single-ion anisotropy
(if the latter was appropriate in the analysis of some more general magnetic model), the preferential population
of the $y^{\prime}z^{\prime}$ and $z^{\prime}x^{\prime}$ orbitals typically stabilizes
the out-of-plane configuration of spin and orbital magnetic moments. However, it should not be confused with the
present situation, where we deal with the \textit{intersite} interactions, which are governed by completely different
processes rather than the single-ion anisotropy energy.
This example emphasizes the importance of the tetragonal crystal-field splitting, which is sometimes ignored
during the construction of the pseudospin models.\cite{Igarashi}

  Due to the tetragonal $I4/mmm$ symmetry, the g-tensor of Ba$_2$IrO$_4$ has
only two inequivalent matrix elements: $g^{xx} = g^{yy}$ and $g^{zz}$. Other
elements are identically equal to zero. The value of $g^{xx}$ and $g^{zz}$
are listed in Table~\ref{tab:g-tensor} together with the partial
contributions of the spin and orbital components.

\begin{table}[h!]
\caption{Matrix elements of the g-tensor, obtained in the six-orbital model
for Ba$_2$IrO$_4$ and Sr$_2$IrO$_4$, and results of their decomposition into
the spin ($S$) and orbital ($L$) parts (given in parenthesis).}
\label{tab:g-tensor}
\begin{ruledtabular}
\begin{tabular}{cccc}
  Compound                        & $g^{xx}$ $(g^{xx}_S,g^{xx}_L)$ & $g^{zz}$ $(g^{zz}_S,g^{zz}_L)$ & $g^{xy}$ $(g^{xy}_S,g^{xy}_L)$ \\
\hline
Ba$_2$IrO$_4$ & $1.796$ $(0.545,1.251)$ & $2.380$ $(0.909,1.470)$ & $0$ $(0,0)$ \\
Sr$_2$IrO$_4$ & $1.115$ $(0.208,0.907)$ & $3.332$ $(1.582,1.750)$ & $0.005$ $(0,0.005)$ \\
\end{tabular}
\end{ruledtabular}
\end{table}

\subsection{\label{sec:6SIO} Six-orbital model for Sr$_2$IrO$_4$}
In the case of Sr$_2$IrO$_4$, the splitting of the $t_{2g}$ levels
is $-$$431$, $-$$4$, and $435$ meV. The symmetry properties of the
$|$$+$$z$$\rangle$ orbital are
given by the same Eq.~(\ref{eqn:z+}) with the following (nonvanishing) coefficients:
$c^{\downarrow}_{xy} = -0.015 \mp i 0.087$,
$c^{\uparrow}_{zx} = -$$i c^{\uparrow}_{yz} = \pm 0.184 + i0.643$, and
$c^{\downarrow}_{x^2-y^2} = \pm 0.004 - i 0.311$,
where the upper (lower) sign is referred to the site $1$ ($2$), experiencing the
counterclockwise (clockwise) rotation of the IrO$_6$ octahedra (see Fig.~\ref{fig.structure}).

  In the local representation, which diagonalizes the
site-diagonal part $[t^{\alpha \beta}_{i = j}]$ of the one-electron Hamiltonian,
the matrix of transfer integrals between sites $1$ and $2$ in the $xy$
planes is given by (in meV)
\begin{equation*}
\hat{t}_{\langle ij \rangle || x^{\prime},y^{\prime}} = \left(
\begin{array}{cccccc}
218 + i60 & 0 & 0 & \mp 24 \mp i11 & 0 & 4 - i38 \\
0 & 218 -i60 & \pm 24 \mp i11 & 0 & -4 - i38 & 0 \\
0 & \mp 24 \pm i11 & -94 -i69 & 0 & \pm 29 \mp i68 & 0 \\
\pm 24 \pm i11 & 0 & 0 & -94 +i69 & 0 & \pm 29 \pm i68 \\
0 & -4 - i38 & \mp 29 \pm i68 & 0 & -144 -i7 & 0 \\
4 - i38 & 0 & 0 & \mp 29 \mp i68 & 0 & -144 + i7
\end{array}
\right),
\end{equation*}
where the upper (lower) sign stands for the bond parallel to the $x^{\prime}$
($y^{\prime}$) axis (see Fig.~\ref{fig.structure} for the notations). This
matrix has both hermitian $\hat{t}_{ij}^{\mathrm{h}} = \frac{1}{2}(\hat{t}_{ij} + \hat{t}_{ji})$ and antihermitian $\hat{t}_{ij}^{\mathrm{ah}} = \frac{1}{2}(\hat{t}_{ij} - \hat{t}_{ji})$ parts. The hermitian part has the same
form as in Ba$_2$IrO$_4$, where the off-diagonal matrix elements give rise
to symmetric anisotropic interactions $\tensor{\boldsymbol{J}}_{ij}^{(S)}$. The
alternation of signs of some of these matrix elements will also lead to the
alternation of anisotropic interactions in the $xy$ plane. The
antihermitian part is the new aspect, which is related to the fact that
the neighboring Ir sites in the $I4_1/acd$ structure are no longer
connected by the inversion operation. This part is responsible for the DM
interactions. The transfer integrals, involving the highest Kramer's doublet are
generally smaller in Sr$_2$IrO$_4$ in comparison with Ba$_2$IrO$_4$,
mainly due to the additional rotation of the IrO$_6$ octahedra and deformation
of the Ir-O-Ir bonds. Therefore, the SE interactions are also expected to be smaller
in Sr$_2$IrO$_4$.

  Due to the additional symmetry lowering, the matrix of the screened Coulomb
interactions $[ U_{\alpha \beta \gamma \delta} ]$ is even
more complex than in Ba$_2$IrO$_4$. Nevertheless, the energies
of the two-hole states, obtained from $[ U_{\alpha \beta \gamma \delta} ]$,
have the same ``three-level'' structure as in Ba$_2$IrO$_4$, which is only slightly
deformed by the lattice distortion and the SO interaction (see Fig.~\ref{fig.multiplet}).
The averaged parameters $\mathcal{U}$ and $\mathcal{J}$
can be again evaluated from the splitting between these three groups of levels
as $\mathcal{U} = 3.05$ eV and $\mathcal{J} = 0.48$ eV
(both with and without the SO interaction). The value of $\mathcal{J}$ is comparable with the
one in Ba$_2$IrO$_4$. However, the Coulomb repulsion $\mathcal{U}$ is slightly larger in Sr$_2$IrO$_4$.
This behavior is consistent with the change of the electronic structure
(see Figs.~\ref{fig.BIODOS} and \ref{fig.SIODOS}): since the unoccupied Ba $5d$ states are closer
to the Fermi level and strongly hybridize with the Ir $5d$ states, the Coulomb $\mathcal{U}$
is expected to be more screened in Ba$_2$IrO$_4$ than in Sr$_2$IrO$_4$.\cite{review2008}
Moreover, it is reasonable to expect that the additional $I4_1/acd$ distortion in the case of Sr$_2$IrO$_4$
will make the $t_{2g}$ states more localized and, thus, the screening of $\mathcal{U}$ less efficient.
This will further reduce the values of the SE interactions in Sr$_2$IrO$_4$.

  Considering only the values of interorbital Coulomb interactions
$\mathcal{U}'=\mathcal{U} - 2\mathcal{J} = 1.90$ eV and $2.09$ eV for Ba$_2$IrO$_4$ and Sr$_2$IrO$_4$, respectively,
we note a reasonable agreement with the results full-scale constrained RPA calculations
reported in Ref.~\onlinecite{Arita} ($\mathcal{U}'$ is about $1.47$ eV and $1.77$ eV for Ba$_2$IrO$_4$ and Sr$_2$IrO$_4$, respectively).
Moreover,
the authors of Ref.~\onlinecite{Arita} used a simplified $I4/mmm$ structure and theoretical lattice parameters both for
Ba$_2$IrO$_4$ and Sr$_2$IrO$_4$, which may lead to the
additional screening of $\mathcal{U}'$.
A more serious discrepancy is found for $\mathcal{J}$: our value of
$\mathcal{J}$ is close to the atomic one,
which seems to be reasonable, because $\mathcal{J}$ is only weakly screened in RPA.\cite{PRL05}
However, the values of $\mathcal{J}$ reported in
Ref.~\onlinecite{Arita} are about three times smaller, leading to the violation of the Kanamori rule
$\mathcal{U}'=\mathcal{U} - 2\mathcal{J}$, presumably due to the contribution of the oxygen states to the
Wannier functions.\cite{UW}
This itself is an interesting point, because, according to Ref.~\onlinecite{Khaliullin}, smaller value of $\mathcal{J}$
within the spherical model, which respects the Kanamori rule, should reduce
the anisotropy of the exchange interactions. Therefore, it is interesting
to which extent this anisotropy of the exchange interactions will be compensated
by the anisotropy of the Coulomb interactions, which emerges in the full-scale constrained RPA calculations and manifested
in the violation of the Kanamori rule.
In any case, according to the analysis of the
effective electron model based on the dynamical mean-field theory,\cite{Arita}
our values of the parameters $\mathcal{U}$ and $\mathcal{U}'$ should correspond to the insulating behavior for
Ba$_2$IrO$_4$ and Sr$_2$IrO$_4$, thus justifying the use of the $1/\mathcal{U}$ expansion for the analysis of
the exchange interactions.

  The $I4_1/acd$ structure of Sr$_2$IrO$_4$ contains two IrO$_2$ planes.
The behavior of NN SE interactions in one of the plane is explained in Fig.~\ref{fig.6oPlane}.
The parameters in another plane can be obtained by the $90^\circ$-rotation about the $z$-axis.
As was expected, the isotropic part of the exchange interactions
$J_{12} = \frac{1}{3}(J^{xx}_{12}$$+$$J^{yy}_{12}$$+$$J^{zz}_{12}) = 39.0$ meV is
considerably smaller than in Ba$_2$IrO$_4$.

  Since $J^{xx}_{12} = J^{yy}_{12} > J^{zz}_{12}$,
the pseudospins will favor the in-plane configuration, similar
to Ba$_2$IrO$_4$ and in agreement with the experimental situation.\cite{BJKim,Ye}
In Sr$_2$IrO$_4$, the parameter of the easy-plane anisotropy for the NN interactions,
$\Delta_{\lambda} = 1 - J^{zz}_{12}/J^{xx}_{12}$,
has been recently estimated in the X-ray resonant magnetic scattering experiments as $0.08$,\cite{Des}
which is close to our theoretical value of $0.087$.
The symmetric anisotropic part of $\tensor{\boldsymbol{J}}_{12}$
is $|J^{(S)xy}_{12}| \equiv \Delta J_{12} = 0.73$ meV, which is about one order of magnitude
smaller than in Ba$_2$IrO$_4$. This interaction is also bond-dependent.

  The antisymmetric part of $\tensor{\boldsymbol{J}}_{12}$ can be represented in terms of the of
DM vector (in meV): $\boldsymbol{d}_{12} = (-$$0.1,-$$0.1,-$$3.97)$
(see Fig.~\ref{fig.structure} for the notations of the atomic sites).
The phases of $d^x$ and $d^y$ alternate in the four NN bonds around the site $1$.
Therefore, since all NN atoms, surrounding the site $1$, have the same direction of the
pseudospin, the total contribution of $d^x$ and $d^y$ to the canting of these pseudospins
will vanish. On the other hand, the phases of $d^z$ are the same for all NN bonds.
Thus, $d^z$ will be responsible for the ferromagnetic (FM) canting, which can be estimated as
$| d^z_{12}/(2 J_{12}^{xx}) | \sim 2.8^\circ$. This value is smaller than the
experimental estimate of $8^\circ$.\cite{Cao}
Nevertheless, the negative sign of $d^z$ for the bond 1-2 is consistent with the counterclockwise
rotation of the IrO$_6$ octahedra.\cite{Khaliullin}
This picture can be also verified experimentally.\cite{Dmitrienko}

  The g-tensor relates the pseudospins with the value of true magnetic moments, which can be
observed in the experiment. Using the value of $g^{xx} = 1.115$ (Table~\ref{tab:g-tensor}),
the local magnetic moment in the $xy$ plane can be estimated as $\frac{1}{2}g^{xx} = 0.56$ $\mu_{\rm B}$,
where the spin and orbital counterparts are
$\frac{1}{2}g^{xx}_S = 0.10$ $\mu_{\rm B}$ and $\frac{1}{2}g^{xx}_L = 0.46$ $\mu_{\rm B}$,
respectively.

\subsection{\label{sec:Neel}Calculations of N\'eel temperature}
Thus, the first-principles calculations have revealed a
big difference of the magnetic models
in the case of Ba$_{2}$IrO$_{4}$ and Sr$_{2}$IrO$_{4}$.
On the one hand, the leading isotropic exchange interaction of $123$ meV in Ba$_{2}$IrO$_{4}$
is about three times larger than that of $39$ meV in Sr$_{2}$IrO$_{4}$.
In turn, the symmetric anisotropic interaction
$J^{(S)xy}_{12}$ in Ba$_2$IrO$_4$ is an order of magnitude larger than in Sr$_{2}$IrO$_{4}$.
On the other hand, there is an appreciable DM interaction in Sr$_{2}$IrO$_{4}$, but not in Ba$_{2}$IrO$_{4}$.
One of the puzzling points is that the experimental
N\'eel temperature remains practically the same in both compounds (about 240 K).
The aim of this section is to check whether such striking similarity can be explained using above
parameters of interatomic exchange interactions derived in the SE approximation.

  Let us first investigate the effect of the DM interaction on the energy spectrum of the pseudospin model.
In the $4m/mmm$ coordinate frame, the exchange interaction tensor in the bond 1-2, which is parallel to the
$y^\prime$ axis, is given by
\begin{equation*}
\tensor{\boldsymbol{J}}_{12} =
\left(
\begin{array}{ccc}
J_{12}^{xx} - \Delta J_{12} & -d_{12}^z & 0 \\
\phantom{-}d_{12}^z & J_{12}^{xx} + \Delta J_{12}  & 0 \\
0 & 0 & J_{12}^{zz}
\end{array}
\right),
\end{equation*}
where for simplicity we have dropped the small contributions of $d_{12}^x$ and $d_{12}^y$.
For the bonds parallel to the $x^\prime$ axis, $\Delta J_{12}$ should be replaced by $-$$\Delta J_{12}$.
By considering the transformation,
\begin{equation*}
\tensor{\boldsymbol{J}}_{12} \rightarrow \tensor{\tilde{\boldsymbol{J}}}_{12} = \tensor{U}_1 \tensor{\boldsymbol{J}}_{12} \tensor{U}_2^T
\end{equation*}
with
\begin{equation*}
\tensor{U}_1 = \tensor{U}_2^T =
\left(
\begin{array}{ccc}
\phantom{-}\cos \phi & \phantom{-}\sin \phi & 0 \\
-\sin \phi & \phantom{-}\cos \phi  & 0 \\
0 & 0 & 1
\end{array}
\right)
\end{equation*}
and $\phi = \frac{1}{2} \arctan(d_{12}^z/J_{12}^{xx})$ minimizes the energy of DM interactions,\cite{Khaliullin,Aharony}
the tensor $\tensor{\boldsymbol{J}}_{12}$ can be transformed to
\begin{equation*}
\tensor{\tilde{\boldsymbol{J}}}_{12} =
\left(
\begin{array}{ccc}
\tilde{J}_{12}^{xx} - \Delta J_{12} & 0 & 0 \\
0 & \tilde{J}_{12}^{xx} + \Delta J_{12}  & 0 \\
0 & 0 & J_{12}^{zz}
\end{array}
\right),
\end{equation*}
where $\tilde{J}_{12}^{xx} = J_{12}^{xx} \sqrt{1 + (d_{12}^z/J_{12}^{xx})^2}$. Thus, the DM interactions alone
do not confine the pseudospins in any particular directions.\cite{Khaliullin,Aharony}
Moreover, after such transformation to the local coordinate frame, the effect of the DM interactions can be combined with $J_{12}^{xx}$.
Since in the 6-orbital model for Sr$_2$IrO$_4$, $d^z_{12} = 3.97$ meV while $J_{12}^{xx} = 40.2$ meV, the
renormalization of $\tilde{J}_{12}^{xx}$ due to the DM interaction is only about $0.5$~\%.
Therefore, we conclude that the effect of the DM interaction on the energy spectrum
is small and can be neglected and, as far as the energy spectrum is concerned, the main ingredients of the pseudospin model
are essentially the same in the case of Ba$_2$IrO$_4$ and Sr$_2$IrO$_4$.

  Below, we will concentrate on two mechanisms of the magnetic ordering in iridates:
the first one is due to the in-plane anisotropy, which emerges in the 6-orbital model, and the
second one is due to the interlayer exchange coupling, which is relevant to the 2-orbital model of Ba$_2$IrO$_4$.
Thus, we consider the following general compass Heisenberg model
\begin{eqnarray}
\hat{\mathcal{H}}_{\mathcal{S}} &=&\frac{J_{z}}{2}\sum_{<ij>\text{ in plane}}
\mathcal{S}_{i}^{z}\mathcal{S}_{j}^{z}+
\frac{1}{2}\sum_{<ij>\parallel x}\left( J_{\parallel }\mathcal{S}_{i}^{x}\mathcal{S}_{j}^{x}+
J_{\perp}\mathcal{S}_{i}^{y}\mathcal{S}_{j}^{y}\right)\notag \\
&+&\frac{1}{2}\sum_{<ij>\parallel y}\left(
J_{\parallel }\mathcal{S}_{i}^{y}\mathcal{S}_{j}^{y}+J_{\perp }\mathcal{S}_{i}^{x}\mathcal{S}_{j}^{x}\right)
+\frac{J^{\prime }}{2}\sum_{<ij>\text{ inter plane}}\boldsymbol{ \cal S}_{i}\boldsymbol{ \cal S}_{j}, \label{eqn:Ham}
\end{eqnarray}
where it is convenient to introduce the shorthand notations:
$J_z \equiv J_{12}^{zz}$, $J_{\parallel } \equiv J_{12}^{xx} + \Delta J_{12}$,
$J_{\perp } \equiv J_{12}^{xx} - \Delta J_{12}$, and
$J^{\prime }$ is the coupling between the atoms, which belong to different planes, separated by the primitive
translation $c$ along the $z$ axis.
The magnon spectrum of this model for $J^{\prime }=0$ was calculated in Ref.~\onlinecite{Yildirim}. It reads
\begin{eqnarray}
E_{\mathbf{q}}^{(1)} &=& \zeta S\sqrt{(4J_{\mathrm{av}}+B_{q}-A_{q})(4J_{\mathrm{av}}+B_{q}+A_{q}+J_{\mathrm{av}}g)}, \notag \\
E_{\mathbf{q}}^{(2)} &=& \zeta S\sqrt{(4J_{\mathrm{av}}-B_{q}-A_{q}+J_{\mathrm{av}}g)(4J_{\mathrm{av}}-B_{q}+A_{q})},
\end{eqnarray}
where $J_{\mathrm{av}}=(J_{\parallel }+J_{\perp })/2=J_{12}^{xx}$,
\begin{eqnarray}
A_{\mathbf{q}} &=&(J_{\perp }+J_{z})\cos q_{x}+(J_{\parallel }+J_{z})\cos
q_{y}, \notag \\
B_{\mathbf{q}} &=&(J_{\perp }-J_{z})\cos q_{x}+(J_{\parallel }-J_{z})\cos
q_{y},
\end{eqnarray}
and
\begin{equation}
g=0.16(J_{\parallel }-J_{\perp })^{2}/(J_{\mathrm{av}}^{2} S)=0.64(\Delta J_{12}/J_{12}^{xx})^{2}/S \label{qg}
\end{equation}
is the quantum gap. Moreover, we have introduced the renormalization factor $\zeta =1+0.0785/S$, which
is taken equal to its value in the two-dimensional Heisenberg model. Then, for small $\mathbf{q}$ we obtain
\begin{eqnarray}
4J_{\mathrm{av}}-B_{\mathbf{q}}-A_{\mathbf{q}} &\rightarrow &J_{\perp
}q_{x}^{2}+J_{\parallel }q_{y}^{2}, \notag \\
4J_{\mathrm{av}}-B_{\mathbf{q}}+A_{\mathbf{q}} &\rightarrow &4(J_{\mathrm{av}
}+J_{z}), \notag \\
4J_{\mathrm{av}}+B_{\mathbf{q}}-A_{\mathbf{q}} &\rightarrow &4(J_{\mathrm{av}
}-J_{z})+J_{z}q^{2}=J_{z}\left( q^{2}+f\right), \notag \\
4J_{\mathrm{av}}+B_{q}+A_{q} &\rightarrow &8J_{\mathrm{av}},
\end{eqnarray}
where the parameter $f$ describing the in-plane symmetric anisotropy is defined as
\begin{equation}
f=4(J_{\mathrm{av}}-J_{z})/J_{z}  \label{f}.
\end{equation}
Therefore, we have:
\begin{eqnarray}
E_{\mathbf{q}}^{(1)} &\simeq &S \zeta \sqrt{8J_{\mathrm{av}}J_{z}\left(
q^{2}+f\right) }, \notag \\
E_{\mathbf{q}}^{(2)} &\simeq &S \zeta \sqrt{4(J_{\mathrm{av}}+J_{z})(J_{\perp}q_{x}^{2}+J_{\parallel }q_{y}^{2}+J_{\mathrm{av}}g)}. \label{cont:Spec}
\end{eqnarray}
The first mode is related to the out-of-plane pseudospin rotation, while the
second corresponds to the in-plane rotation.

  To obtain magnetic transition temperatures, we map the Heisenberg model (\ref{eqn:Ham})
onto the non-linear sigma model, having the same
excitation spectrum, Eq. (\ref{cont:Spec}), see Appendix. Treating the
magnetic excitations, as slightly different from the case of the $XY$ anisotropy,\cite{Katanin}
we obtain in the regime $f\gg \max(\alpha,g)$, $\alpha =2J^{\prime }/J$,
the following equation for the Neel temperature (see Appendix):
\begin{equation}
T_{N}=4\pi \rho _{s}\left\{ \ln \frac{T_{N}^{2}}{c_{\mathrm{op}}c_{\mathrm{ip}
}f_{r}}+4\ln \frac{4\pi \rho _{s}}{T_{N}}-\frac{2A^{2}}{\ln ^{2}\left(
f/\max (\alpha,g) \right) }\right\} ^{-1},  \label{tc}
\end{equation}
where $A\simeq 3.5$, $c_{\mathrm{op}}=\sqrt{8J_{\mathrm{av}}J_{z}}S\zeta $ and
$c_{\mathrm{ip}}=\sqrt{4J_{\mathrm{av}}(J_{\mathrm{av}}+J_{z})}S\zeta$ are the
out-of-plane and in-plane spin-wave velocities, $\rho
_{s}=2(1/\rho _{\mathrm{z}}+1/\rho _{\mathrm{av}})^{-1}$ is the effective spin stiffness
($\rho _{\mathrm{
z,av}}=J_{z,\mathrm{av}}\zeta S\overline{S}_{0}$), $f_{r}=f(\overline{S}
_{0}/S)^{2}$ is the renormalized anisotropy parameter, $\overline{S}_{0}=0.303$ for $S=1/2$ is the
ground-state magnetization.
In the absence of compass anisotropy, $f=g=0$, we obtain
instead \cite{Katanin1}
\begin{equation}
T_{N}=4\pi \rho _{s}\left\{ \ln \frac{2T_{N}^{2}}{c_{\mathrm{op}}c_{\mathrm{ip}
}\alpha _{r}}+3\ln \frac{4\pi \rho _{s}}{T_{N}}-0.06\right\} ^{-1}, \label{tc1}
\end{equation}
where $\alpha _{r}=\alpha (\overline{S}_{0}/S)$ is the renormalized interlayer coupling parameter.

  The parameters and the resulting magnetic transition temperatures are listed in Table \ref{betaJ}.
\begin{table}[!h]
\centering
\caption{Parameters used in Eqs.~(\ref{tc}) and (\ref{tc1}) for the transition temperature equation
and the calculated $T_{N}$ in different regimes (the values of $J_{\mathrm{av}}$, $J_{\parallel }-J_{\perp }$, and
$J_{z}$ are in meV, $T_{N}$ is in Kelvins, and other parameters are dimensionless).}
\label{betaJ}
\begin{tabular}{ccccccccccc}
\hline
& $J_{\mathrm{av}}$ & $J_{\parallel }-J_{\perp }$ & $J_{z}$ & $f$ & $g$ & $
\alpha $ & $c_{\mathrm{op}}$ & $c_{\mathrm{ip}}$ & $T_{N}^{\alpha =0, g=0}$ &
$T_{N}$  \\ \hline
Ba$_{2}$IrO$_{4}$ (2-orb.) & 122.8 & 0 & 122.8 & 0 & 0 & 1.4$\cdot$10$^{-4}$ & 200.9 & 200.9 & -  & 239  \\ \hline
Ba$_{2}$IrO$_{4}$ (6-orb.) & 110.6 & 15 & 108.1 & 0.09 & 6$\cdot $10$^{-3}$ & 1.3$\cdot$10$^{-4}$ & 178.9 & 179.9 & 371 & 414  \\ \hline
Sr$_{2}$IrO$_{4}$ (6-orb.) & 40.2 & 1.5 & 36.7 & 0.38 & 4$\cdot $10$^{-4}$ & 1.5$\cdot$10$^{-5}$ & 62.8 & 64.3 & 181 & 216  \\ \hline
\end{tabular}
\end{table}
Lets us first discuss the results of the 6-orbital models for the Ba$_{2}$IrO$_{4}$ and Sr$_{2}$IrO$_4$.
Judging by the ratio between the anisotropy parameters $f,g$ and interlayer isotropic parameter, $\alpha$,
we have the relation $f\gg g\gg \alpha$,
which holds for both compounds.
Thus, the in-plane anisotropy is expected to be mainly responsible
for the magnetic ordering.
The differences between in-plane and out-of-plane components of the symmetric anisotropy tensor,  ($J_{\mathrm{av}} - J_z$), are close to each other and equal to 2.5 meV (in Ba$_2$IrO$_4$) and 3.5 meV (in Sr$_2$IrO$_4$). However, due to the difference in
the absolute value of $J_{z}$, we obtain completely different anisotropy parameters $f,g$ and,
therefore, the transition temperatures.
For Sr$_2$IrO$_4$, the calculated temperature of 216 K is in
the good agreement with the experimental value of 240 K.
This is consistent with the finding of Jackeli and Khaliullin (Ref.~\onlinecite{Khaliullin}), who used the experimental
$T_{N}$ in order to estimate the values of the exchange interactions and these values are close to ours.
However, the situation is different in the case of Ba$_2$IrO$_4$, where
the theoretical $T_{N}$ is overestimates by factor two.
Interestingly, in the case of the 2-orbital model for Ba$_2$IrO$_4$, which, in analogy with the cuprates,\cite{Katanin1} contains only in-plane and inter-plane isotropic exchange interactions, we observe a good agreement between theory and experiment. However, this agreement is probably fortuitous.

\section{\label{sec:beyondSE}Beyond Superexchange}
The main purpose of this section is to discuss the effect, which are not included to the regular SE model.
Our main concern is the following: since the SE model is based on the second-order perturbation theory for the
transfer integrals, it implies that all effects of the SO coupling, which are included to these transfer integrals,
are also automatically treated only up to the second order. Since the SO coupling is large in iridates, this may be
rather crude approximation, which does not take into account some important anisotropic interactions.
For instance, in the mean-field approximation for the SE model,
all pseudospins in the single $xy$ plane of Ba$_2$IrO$_4$ and Sr$_2$IrO$_4$
can rotate rigidly at no energy cost. Besides quantum effects, considered in the previous section
(see also Ref.~\onlinecite{KatukuriPRX}),
this maybe related to the lack of the in-plane anisotropy, which typically
appears only in the fourth order of the SO coupling.

  If we wanted to include these effects in the pseudospin model (\ref{eqn:HS}), our strategy would be to go beyond
the second order perturbation theory for the transfer integrals and consider higher-order terms, which give rise
to the new type of interactions, such as biquadratic and ring exchange.\cite{Nagaev,KataninKampf} They will affect
both anisotropic and isotropic parts of the exchange interactions. Therefore, in the pseudospin formulation, based on the
strong SO coupling, these two types of the effects are connected with each other:
if we want to consider the higher order anisotropic interactions, we have to deal with the biquadratic and ring exchange terms,
which will affect all other exchange interactions, including the isotropic ones.
Such pseudopsin Hamiltonian is no longer presented in the bilinear form (\ref{eqn:HS}).

  Nevertheless, in the present work we take a different strategy and
in order to evaluate the higher-order contributions
(and, therefore, check the validity of the SE model) in Ba$_2$IrO$_4$ and Sr$_2$IrO$_4$,
we solve the original electron model (\ref{eqn:Hel}) in the mean-field
HF approximation, where we also apply the staggered external magnetic field, which controls the directions of
the spin and orbital moments. We have found that the field of $\mu_{\rm B}H=0.68$ meV is generally sufficient for these purposes.

  The weak point of the HF approach is that it treats all on-site
electron-electron interactions on the mean-field level, whereas in the SE theory
such processes are treated rigorously by solving the exact eigenstates problem for the virtual two-hole states.
However, in this particular case, we do not expect large error caused by the mean-field approximation
(some comparison for transition-metal perovskite oxides can be found in Ref.~\onlinecite{review2008}).
On the other hand, the HF method does not employ any additional approximations regarding the relative strength of
transfer integrals and the on-site Coulomb repulsion and, in this sense, is the more superior approach in comparison
with the SE theory.

\begin{figure}[tbp]
\begin{center}
\includegraphics[height=7cm]{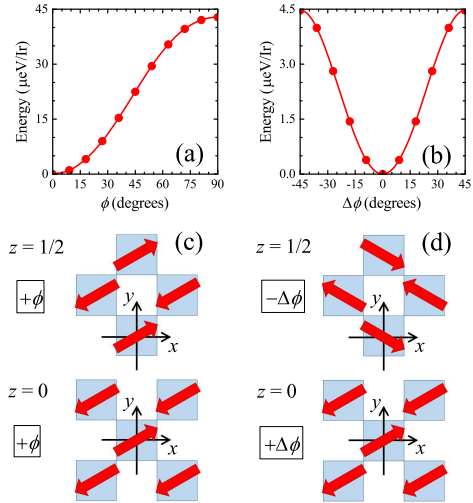}
\end{center}
\caption{(Color online) Results of constrained unrestricted Hartree-Fock calculations for Ba$_2$IrO$_4$
in the staggered ``antiferromagnetic'' field $\mu_{\rm B}H=0.68$ meV.
The direction of the magnetic field in the planes $z=0$ and
$z=1/2$ is specified by azimuthal angles $(\phi$$+$$\Delta \phi)$ and $(\phi$$-$$\Delta \phi)$, respectively
(in the $I4_1/acd$ coordinate frame). (a) is the total energy dependence on $\phi$ for $\Delta \phi = 0$.
(b) is the total energy dependence on $\Delta \phi$ for $\phi = 0$. (c) and (d) explains the geometry of
the staggered magnetic field for (a) and (b), respectively.}
\label{fig.BaH}
\end{figure}

\begin{figure}[tbp]
\begin{center}
\includegraphics[height=7cm]{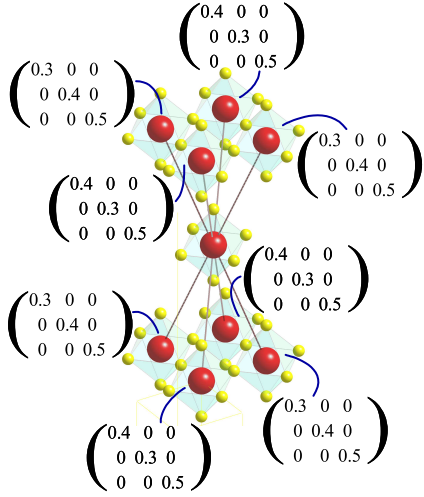}
\end{center}
\caption{(Color online) Tensors of superexchange interactions
$\tensor{\boldsymbol{J}}_{ij}$ (in meV and in the $I4_1/acd$ coordinate frame),
associated with different Ir-Ir bonds between adjacent planes in Ba$_2$IrO$_4$.
The Ir atoms are indicated by the big (red) spheres and the oxygen atoms are
indicated by the small (yellow) spheres.}
\label{fig.6oBBetween}
\end{figure}

  Let us start with Ba$_2$IrO$_4$. The geometry of the constraining field in this case is explained in Fig.~\ref{fig.BaH}.
First, let us consider the case, where the fields in the two adjacent planes $z=0$ and $z=1/2$ are rotated in phase.
Then, the total energy exhibits the minimum at $\phi=0$ (modulo $\pi$, in the $I4_1/acd$ coordinate frame).
This effect can be actually included in the SE model and is related to the anisotropy of the exchange interactions
between adjacent planes.\cite{KatukuriPRX}
The behavior of these interactions is explained in Fig.~\ref{fig.6oBBetween}.
Then, the mean-field energy of the magnetic order, depicted in Fig.~\ref{fig.BaH}c, is given by
$E(\phi) = - \Delta J_{\rm out} \cos 2\phi$ (per one Ir site),
where $\Delta J_{\rm out} = |J_{\rm out}^{xx} - J_{\rm out}^{yy}|$.

  Thus, in the SE approximation, the energy should remain invariant with respect to the antiphase
rotations of the pseudospin (Fig.~\ref{fig.BaH}d). In the other words, if we fix $\phi$ and consider
the configurations, where the directions of the pseudospins in the adjacent planes $z=0$ and $z=1/2$
are specified by the azimuthal angles $(\phi$$+$$\Delta \phi)$ and $(\phi$$-$$\Delta \phi)$, respectively,
the mean-field energy of such configurations should not depend on $\Delta \phi$.\cite{KatukuriPRX}
This property is indeed strictly observed when we use the exchange parameters, derived in the SE model.
Because of this degeneracy, the authors of Ref.~\onlinecite{KatukuriPRX} had to go beyond the mean-field
theory and consider the effect of the quantum fluctuations in order to explain the experimentally observed
magnetic ground-state structure of Ba$_2$IrO$_4$ (corresponding to $\phi = \Delta \phi = 0$ in the $I4_1/acd$ coordinate frame).\cite{Boseggia}
The most interesting aspect of our analysis is that this degeneracy can be lifted even on the mean-field level
if one goes beyond the SE model and consider more rigorously the higher-order contributions of the
transfer integrals in the framework of the unrestricted HF calculations.
The dependence of the HF total energy on $\Delta \phi$ is shown in Fig.~\ref{fig.6oBBetween} (for $\phi=0$).
It clearly shows that the higher order-anisotropic interactions, which are included in the
HF calculations, lifts the degeneracy and stabilizes the experimentally observed magnetic
ground state. The energy barrier, caused by these interactions, is about $4.5$ $\mu$eV, which is at least
comparable with the effect of quantum fluctuations considered in Ref.~\onlinecite{KatukuriPRX}.
Thus, the effect is robust and cannot be neglected in the realistic analysis of the magnetic properties of Ba$_2$IrO$_4$.

  Next, we evaluate the effect of biquadratic exchange on the NN interaction $J_{12}^{zz}$ in the $xy$-plane of Ba$_2$IrO$_4$.
If the magnetic properties of some material were indeed described by the bilinear Hamiltonian (\ref{eqn:HS}),
the values of the exchange parameters would not depend on the method, which is used for their calculations.
For instance, in the mean-field HF method, one could evaluate $J_{12}^{zz}$ from the total energy difference
between FM and AFM states, by aligning the magnetic moments parallel to the $z$ axis:
$J_{12}^{zz} = E_{\uparrow \uparrow}$$-$$E_{\uparrow \downarrow}$. Then, if the bilinear parametrization (\ref{eqn:HS})
for the magnetic Hamiltonian were indeed appropriate,
this value of $J_{12}^{zz}$ should be close
to the one obtained in the SE model. Nevertheless, the straightforward HF calculations yield
$E_{\uparrow \uparrow}$$-$$E_{\uparrow \downarrow} = 83.8$ meV, which is 22\% smaller than
$J_{12}^{zz} = 108.1$ meV, obtained in the SE model. This deviation is the measure of biquadratic
(or ring-type) exchange interactions, existing in the system.
Thus, as expected from the discussion in the beginning of this Section, the higher order anisotropic
effects in Ba$_2$IrO$_4$ coexist with appreciable biquadratic contributions to the isotropic exchange interaction.
In this sense, we obtain very consistent
description for Ba$_2$IrO$_4$. Unfortunately, we could not obtain a stable in-plane FM solution in the HF method and, thus,
evaluate the in-plane elements of the exchange tensor from the total energy difference. Generally, one can expect similar
contribution of biquadratic interactions to the in-plane and out-of-plane components
of the exchange tensor.

  The behavior of Sr$_2$IrO$_4$ appears to be rather different from Ba$_2$IrO$_4$. Since the transfer integrals are
smaller in Sr$_2$IrO$_4$, while the Coulomb interactions are slightly larger, it is reasonable to expect that
the $t_{2g}$ states are more localized in Sr$_2$IrO$_4$, which additionally justifies the use of the SE model.
This is indeed what we have obtained by comparing results of
HF calculations and the SE model. The fact that Ba$_2$IrO$_4$ appears to be ``more itinerant'' than Sr$_2$IrO$_4$
can be seen already from the comparison of the band gap, obtained in the HF method for the AFM ground state, which is
substantially smaller in Ba$_2$IrO$_4$ ($1.3$ eV, against $1.8$ eV in Sr$_2$IrO$_4$). It should be noted, however,
that the HF gap is considerably larger than the experimental one, due to the lack of quantum and thermal fluctuations,
as was confirmed by the DMFT calculations.\cite{Arita}

  First, we consider the HF solutions for the FM and AFM states, where all
magnetic moments are parallel to the $z$ axis. The total energy difference between these states is $31.7$ meV,
which is much closer to the value $J_{12}^{zz} = 36.7$ meV, obtained in the SE model (the difference is about 14\%,
which can be again regarded as the measure of biquadratic interactions in the system). In Sr$_2$IrO$_4$, it is practically
impossible to evaluate the in-plane elements of the exchange tensor from the total energy difference: because of the DM interaction,
the in-plane FM state is unstable and converges to the AFM state (with small FM canting of the magnetic moments).

\begin{figure}[tbp]
\begin{center}
\includegraphics[height=8cm]{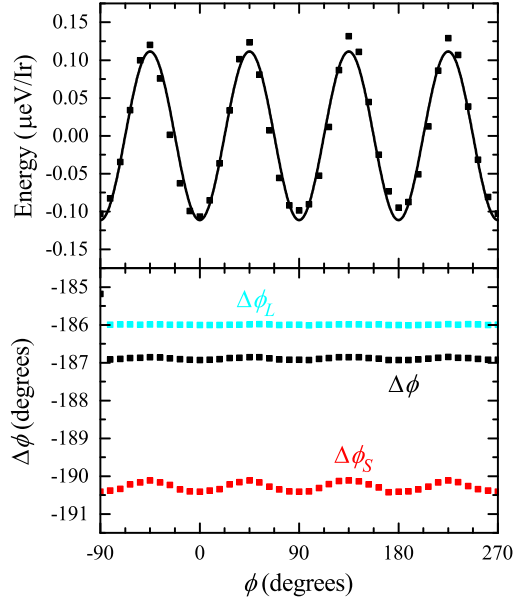}
\end{center}
\caption{Results of constrained unrestricted Hartree-Fock calculations for Sr$_2$IrO$_4$
in the magnetic field $\mu_{\rm B}H=0.68$ meV, which couples to the weak ferromagnetic moment in the $xy$ plane.
The direction of the field is specified by the azimuthal angle $\phi$.
The upper panel displays the behavior of the total energy: the symbols show calculated points, while the solid line is the result of interpolation
$E(\phi) = A + B\cos 4\phi$.
The lower panel shows the angle between spin ($\Delta \phi_S$), orbital ($\Delta \phi_L$) and
total ($\Delta \phi$) magnetic moments of the sites $2$ and $1$
in Fig.~\protect\ref{fig.structure}.}
\label{fig.SrH}
\end{figure}

  Next, we consider the higher-order anisotropy effects in Sr$_2$IrO$_4$. For these purposes we take the
weakly FM state and rotate magnetic moments by the external magnetic field of $\mu_{\rm B}H=0.68$ meV, which
couples to the weak FM moment in the $xy$ plane.
The results of such constrained HF calculations are summarized in Fig.~\ref{fig.SrH}.
We note the following: (i) The total energy depends on the direction of the magnetic moments in the $xy$ plane.
However, this dependence is very weak (the characteristic energy barrier is about $0.25$ meV, which is
an order of magnitude smaller than in Ba$_2$IrO$_4$); (ii) The angle ($\Delta \phi$) between magnetic moments of the sites $2$ and $1$
(see Fig.~\ref{fig.structure} for the notations) is nearly constant, meaning that it is mainly controlled by the DM interaction $d_{12}^z$,
while the effect of other anisotropic interactions (that are not taken into account in the SE model) are relatively small.
Since the energy gain caused by the DM interaction is proportional to $d_{12}^z\sin \Delta \phi$, the obtained values of
$-$$270^\circ < \Delta \phi<-$$180^\circ$ are well consistent with the sign $d_{12}^z < 0$ of DM interactions for the counterclockwise rotation
of the IrO$_6$ octahedra around the site $1$ (see Fig.~\ref{fig.structure}). Yet, one interesting aspect of the HF analysis is that
the angle $\Delta \phi$ is different between, separately, spin and orbital magnetic moments. Without external field ($H=0$),
$\Delta \phi$ is about $-$$185.2^\circ$. It corresponds to the FM canting of $2.6^\circ$, which is close to $2.8^\circ$,
obtained in the SE model. The values of spin and orbital magnetic moments, obtained for the in-plane (out-of-plane) magnetic alignment
are $0.13$ and $0.48$ $\mu_{\rm B}$ ($0.71$ and $0.83$ $\mu_{\rm B}$), respectively, which are in good agreement with the values of
corresponding matrix elements of the g-tensor, reported in Table~\ref{tab:g-tensor} for the SE model.

  Thus, we obtain a very consistent description also for Sr$_2$IrO$_4$: (i) To a good approximation, the magnetic Hamiltonian has the
bilinear form (\ref{eqn:HS}), inherent to the SE model; (ii) The higher-order anisotropy effects, beyond the SE model, are negligibly small.
This makes the main difference from Ba$_2$IrO$_4$, where (i) the deviations from the bilinear form are significant and (ii) the higher-order
anisotropic exchange interactions are important.

\section{\label{sec:conc}Summary and Conclusions}
The main purpose of this work was to critically evaluate the abilities of the SE model for the analysis of
magnetic properties of the layered iridates Ba$_2$IrO$_4$ and Sr$_2$IrO$_4$.
Being based on the first-principles electronic structure calculations with the SO coupling, we have first derived the
effective low-energy electron model for the $t_{2g}$ bands, which are located near the Fermi level and primarily
responsible for the magnetic properties of Ba$_2$IrO$_4$ and Sr$_2$IrO$_4$.
This electron model
was further mapped on the pseudospin model using the theory of SE interactions in the limit of large
on-site Coulomb repulsion. We have clarified the microscopic origin of the
bond-dependent anisotropic exchange interactions, as well as
the antisymmetric DM interactions, caused by the anti-phase rotations of the IrO$_6$ octahedra in Sr$_2$IrO$_4$.
The pseudospin Hamiltonian problem has been solved by means of the non-linear sigma model, that has finally allowed
to evaluate the N\'eel temperature for both considered compounds. We have demonstrated that while for Sr$_2$IrO$_4$
the theoretical N\'eel temperature is in good agreement with the experimental data,
for Ba$_2$IrO$_4$ it is overestimated by factor two.
We have argued that this discrepancy is quite consistent with the limitations of the SE model for Ba$_2$IrO$_4$,
which is the more ``itinerant'' system than Sr$_2$IrO$_4$. Such ``itineracy'' is directly related to the details
of the electronic structure of Ba$_2$IrO$_4$: the lack of rotations of the IrO$_6$ octahedra and the proximity of the
Ba $5d$ states to the Fermi level make the $t_{2g}$ bandwidth increase
and more efficiently screen the Coulomb interactions in this band. Thus, the $\hat{t}/\mathcal{U}$ expansion for the
magnetic energy converges slower and higher-order terms, beyond the SE contributions, start to play an important role.
Since the effect of SO interaction in the SE formulation is included to the transfer integrals, the higher-order terms
automatically improve the description also for the anisotropic exchange interactions. In fact, by solving the low-energy electron
model for Ba$_2$IrO$_4$ in the HF approximation, we were able to reproduce the experimental magnetic ground states structure
of this compound
even on the mean-field level, without invoking to quantum effects.

\textit{Acknowledgements}. We acknowledge fruitful communication with Alexander Tsirlin.
This work is partly supported by the grant of
Russian Science Foundation (project No. 14-12-00306).

\appendix*
\section{Derivation of the non-linear sigma model for compass\
Heisenberg model and its renormalization}

\subsection{Nonlinear-sigma model}
To obtain the action of the continuum model we pass to the coherent state
representation for spin operators and represent the corresponding vectors of
spin directions following the standard procedure
\begin{equation}
\boldsymbol{ \cal S}_{i}=(-1)^{i}S\mathbf{n}_{i}\sqrt{1-(\mathbf{L}_{i}/S)^{2}}+
\mathbf{L}_{i},  \label{Repr}
\end{equation}
where $\mathbf{L}_{i}\cdot \mathbf{n}_{i}=0,$ $\mathbf{n}_{i}^{2}=1,$ and
the fields $\mathbf{n}_{i}$ and $\mathbf{L}_{i}$ represent the staggered and
uniform component. Substituting Eq. (\ref{Repr}) into (\ref{eqn:Ham}) we obtain
the Lagrangian:
\begin{align}
L[\mathbf{n},\mathbf{L}]& =-\frac{J_{z}S^{2}}{2}\sum_{i,\delta
}n_{i}^{z}n_{i+\delta }^{z}-\frac{S^{2}}{2}\sum_{i,\delta _{x}}\left(
J_{\parallel }n_{i}^{x}n_{i+\delta _{x}}^{x}+J_{\perp }n_{i}^{y}n_{i+\delta
_{x}}^{y}\right)  \notag \\
 &-\frac{S^{2}}{2}\sum_{i,\delta _{y}}\left( J_{\parallel
}n_{i}^{y}n_{i+\delta _{y}}^{y}+J_{\perp }n_{i}^{x}n_{i+\delta
_{y}}^{x}\right)   \notag \\
& +\frac{1}{2}\sum_{i,\delta }\left( J_{z}L_{i}^{z}L_{i+\delta
}^{z}+(J_{z}n_{z}^{2}+J_{\parallel }n_{\delta }^{2}+J_{\perp }n_{\widetilde{
\delta }}^{2})\frac{\mathbf{L}_{i}^{2}+\mathbf{L}_{i+\delta }^{2}}{2}\right) \notag \\
&+\frac{1}{2}\sum_{i,\delta _{x}}\left( J_{\parallel }L_{i}^{x}L_{i+\delta
_{x}}^{x}+J_{\perp }L_{i}^{y}L_{i+\delta _{x}}^{y}\right)   \notag \\
& +\frac{1}{2}\sum_{i,\delta _{y}}\left( J_{\parallel }L_{i}^{y}L_{i+\delta
_{y}}^{y}+J_{\perp }L_{i}^{x}L_{i+\delta _{y}}^{x}\right) +i\sum_{i}\mathbf{L
}_{i}\mathbf{\cdot \lbrack n}_{i}\mathbf{\times \partial }_{\tau }\mathbf{n}
_{i}\mathbf{]},
\end{align}
where $n_{\delta _{x,y}}=n_{x,y},n_{\widetilde{\delta }_{x,y}}=n_{y,x},$ and
we keep only terms, which do not vanish and give the leading contribution in
the continuum limit. Expanding
\begin{equation}
\mathbf{n}_{i+\delta }=\mathbf{n}_{i}+(\mathbf{\delta \nabla )n}_{i}+\frac{1
}{2}(\delta ^{a}\delta ^{b}\mathbf{\partial }_{a}\mathbf{\partial }_{b}
\mathbf{)n}_{i}+...
\end{equation}
and similarly for $\mathbf{L}_{i+\delta },$ we obtain:
\begin{align}
L[\mathbf{n},\mathbf{L}]& =\frac{S^{2}}{2}\int d^{2}\mathbf{x}\left[ J_{z}
\mathbf{(\nabla }n_{z}\mathbf{)}^{2}+J_{\parallel }(\partial
_{x}n_{x})^{2}+J_{\parallel }(\partial _{y}n_{y})^{2}+J_{\perp }(\partial
_{x}n_{y})^{2}+J_{\perp }(\partial _{y}n_{x})^{2}+J_{z}fn_{z}^{2}\right]
\notag \\
& +\frac{1}{2}\int d^{2}\mathbf{x}\left[ 2(4J_{z}+(J_{\parallel }+J_{\perp
}-2J_{z})(n_{x}^{2}+n_{y}^{2}))\mathbf{L}_{i}^{2}\mathbf{+}2(J_{\parallel
}+J_{\perp })(L_{x}^{2}+L_{y}^{2})\right]   \notag \\
& +i\mathbf{L\cdot \lbrack n\times \partial }_{\tau }\mathbf{n]},
\end{align}
where we have defined $f$ according to (\ref{f}). Performing the integration
over $\mathbf{L},$ we find
\begin{align}
L[\mathbf{n}]& =\frac{S^{2}}{2}\int d^{2}\mathbf{x}\left[ J_{z}\mathbf{
(\nabla }n_{z}\mathbf{)}^{2}+J_{\parallel }(\partial
_{x}n_{x})^{2}+J_{\parallel }(\partial _{y}n_{y})^{2}+J_{\perp }(\partial
_{x}n_{y})^{2}+J_{\perp }(\partial _{y}n_{x})^{2}+J_{z}fn_{z}^{2}\right]
\notag \\
& +\frac{1}{2}\int d^{2}\mathbf{x}\frac{1}{2(4J_{z}+(J_{\parallel }+J_{\perp
}-2J_{z})(n_{x}^{2}+n_{y}^{2}))}\mathbf{[n\times \partial }_{\tau }\mathbf{n]
}_{z}^{2}  \notag \\
& +\frac{1}{2}\int d^{2}\mathbf{x}\frac{1}{2(4J_{z}+(J_{\parallel }+J_{\perp
}-2J_{z})(1+n_{x}^{2}+n_{y}^{2}))}\mathbf{[n\times \partial }_{\tau }\mathbf{
n]}_{x}^{2}  \notag \\
& +\frac{1}{2}\int d^{2}\mathbf{x}\frac{1}{2(4J_{z}+(J_{\parallel }+J_{\perp
}-2J_{z})(1+n_{x}^{2}+n_{y}^{2}))}\mathbf{[n\times \partial }_{\tau }\mathbf{
n]}_{y}^{2}.  \label{Lagr}
\end{align}
In the following we assume the preferable direction of magnetic order along
the $y$ axis. Representing
\begin{equation}
n_{y}=\sqrt{1-n_{x}^{2}-n_{z}^{2}}  \label{ny}
\end{equation}
and expanding in $n_{x,z}$ we obtain two branches of the magnon spectrum
\begin{align}
E_{z}^{2}& =4S^{2}J_{z}(J_{\parallel }+J_{\perp })\left( q^{2}+f\right),
\notag \\
E_{x}^{2}& =2S^{2}(2J_{z}+J_{\parallel }+J_{\perp })(J_{\parallel
}q_{x}^{2}+J_{\perp }q_{y}^{2}),
\end{align}
which coincides with small $q$ expansion of the results of Sec.~\ref{sec:Neel} and Ref.~\onlinecite{Yildirim}.

\subsection{Perturbation theory}
In the following we concentrate on the classical part of the Lagrangian (\ref{Lagr}),
renormalized by the quantum fluctuations,
\begin{equation}
L_{\mathrm{cl}}[\mathbf{n}]=\frac{1}{2}\int d^{2}\mathbf{x}\left\{ \rho _{r}
\left[ \mathbf{(\nabla }n_{z}\mathbf{)}^{2}+f_{r}n_{z}^{2}\right] +\rho
_{\parallel }\left[(\partial _{x}n_{x})^{2}+(\partial
_{y}n_{y})^{2}\right]+\rho _{\perp }\left[(\partial _{x}n_{y})^{2}+
(\partial _{y}n_{x})^{2}\right]\right\}.   \label{Lcl}
\end{equation}
In Eq. (\ref{Lcl}) we use the quantum-renormalized spin stiffnesses, $\rho
_{r}=J_{z}S\overline{S}_{0}\zeta $ and $\rho _{\parallel ,\perp }=J_{\parallel ,\perp
}S \overline{S}_{0}\zeta ,$ where $\overline{S}_{0}=S-0.197$ is the
ground state magnetization of the square-lattice Heisenberg model, $\zeta
=1+0.157/(2S)$ is the exchange parameter renormalization factor, the bare
spin stiffnesses anisotropy,  and the renormalized easy plane anisotropy $f_{r}=f\overline{S}
_{0}^{2}/(S\zeta )^{2}$. Following the standard procedure \cite{CHN,Katanin}, we
assume that the excitations, described by $L_{\mathrm{cl}}[\mathbf{n}]$ are
cut on the ultriviolet at the momentum $\Lambda _{\mathrm{uv}}=T/c,$ where
$c=\sqrt{8}JS\zeta $ is the renormalized spin-wave velocity; the remaining
(non-universal) contribution of the other part of momentum space yields the
abovementioned quantum renormalizations.

Assuming again the long-range order along the $y$-axis, introducing $\pi=(n_{x},n_{z}),$
and using Eq. (\ref{ny}), we obtain
\begin{eqnarray}
L_{\mathrm{cl}}[\mathbf{n}] & =&\frac{1}{2}\int d^{2}\mathbf{x}\bigg \{
\rho_{r}\left[ \mathbf{(\nabla}\pi_{z}\mathbf{)}^{2}+f_{r}\pi_{z}^{2}\right]
+\rho_{\parallel}(\partial_{x}\pi_{x})^{2}+\rho_{\perp}(\partial_{y}\pi
_{x})^{2}  \label{Lcl1} \\
&+&\left. \frac{\rho_{\parallel}}{1-\pi^{2}}(\pi\partial_{y}\pi)^{2}+\frac{
 \rho_{\perp}}{1-\pi^{2}}(\pi\partial_{x}\pi)^{2} \right\}
+\frac{T}{2}\int d^{2}\mathbf{x}\ln(1-\pi^{2})-h\int d^{2}\mathbf{x}\sqrt{
1-\pi^{2}},  \notag
\end{eqnarray}
where the first term in the second line comes from the integration measure
and in the last term we have introduced external magnetic field along $y$
axis, which will be put to zero in the end. To perform renormalization of
Eq. (\ref{Lcl1}), we decouple the interactions via the Wick theorem
\begin{align}
L_{\mathrm{cl}}[\mathbf{n}] & =\frac{1}{2}\int d^{2}\mathbf{x}\left\{
\rho_{r}\left[ \mathbf{(\nabla}\pi_{z}\mathbf{)}^{2}+f_{r}\pi_{z}^{2}\right]
+\rho_{\parallel}(\partial_{x}\pi_{x})^{2}+\rho_{\perp}(\partial_{y}\pi
_{x})^{2}+\rho_{\parallel}\langle\pi_{a}^{2}\rangle(\partial_{y}\pi_{a})^{2}
\right.  \notag \\
& \left.
+\rho_{\perp}\langle\pi_{a}^{2}\rangle(\partial_{x}\pi_{a})^{2}+\rho_{
\parallel}\langle(\partial_{y}\pi_{a})^{2}\rangle\pi_{a}^{2}+\rho_{\perp}
\langle(\partial_{x}\pi_{a})^{2}\rangle\pi_{a}^{2}\right]  \notag \\
& +\frac{1}{2}\int d^{2}\mathbf{x}\left[ h\pi^{2}+\frac{h}{2}(3\langle
\pi_{x}^{2}\rangle+\langle\pi_{z}^{2}\rangle)\pi_{x}^{2}+\frac{h}{2}
(3\langle\pi_{z}^{2}\rangle+\langle\pi_{x}^{2}\rangle)\pi_{z}^{2}-T\pi ^{2}
\right].
\end{align}
Rescaling the fields
\begin{align*}
\pi_{x} & \rightarrow Z_{x}\pi_{x}, \\
\pi_{z} & \rightarrow Z_{z}\pi_{z},
\end{align*}
we obtain renormalized parameters:
\begin{align}
\rho_{R} & =Z_{z}^{2}\left[ \rho_{r}+\rho_{xy}\langle\pi_{z}^{2}\rangle
\right],  \notag \\
\rho_{\parallel R} & =Z_{x}^{2}\left[ \rho_{\parallel}+\rho_{\perp}\langle
\pi_{x}^{2}\rangle\right],  \notag \\
\rho_{\perp R} & =Z_{x}^{2}\left[ \rho_{\perp}+\rho_{\parallel}\langle
\pi_{x}^{2}\rangle\right],  \notag \\
\rho_{R}f_{R}+h_{R} & =Z_{z}^{2}\left[ h+\rho_{r}f_{r}+T\int\frac{d^{2}q}{
(2\pi)^{2}}\frac{\rho_{\parallel}q_{y}^{2}+\rho_{\perp}q_{x}^{2}}{\rho
_{r}(q^{2}+f_{r})+h}-T+\frac{h}{2}(3\langle\pi_{z}^{2}\rangle+\langle\pi
_{x}^{2}\rangle)\right],  \notag \\
\rho_{xy,R}g_{R}+h_{R} & =Z_{x}^{2}\left[ h+T\int\frac{d^{2}q}{(2\pi)^{2}}
\frac{\rho_{\parallel}q_{y}^{2}+\rho_{\perp}q_{x}^{2}}{\rho_{
\parallel}q_{x}^{2}+\rho_{\perp}q_{y}^{2}+\rho_{xy} g+h}-T+\frac{h}{2}
(3\langle\pi_{x}^{2}\rangle+\langle\pi_{z}^{2}\rangle)\right],   \label{Ren}
\end{align}
where $\rho_{xy}=(\rho_{\parallel}+\rho_{\perp})/2$,
\begin{align}
\langle\pi_{z}{}^{2}\rangle & =T\int\frac{d^{2}q}{(2\pi)^{2}}\frac{1}{
\rho_{r}(q^{2}+f_{r})+h},  \notag \\
\langle\pi_{x}{}^{2}\rangle & =T\int\frac{d^{2}q}{(2\pi)^{2}}\frac{1}{
\rho_{\parallel}q_{x}^{2}+\rho_{\perp}q_{y}^{2}+\rho_{xy} g+h},
\end{align}
and $g_{R}$ is the gap, generated for $\pi_{x}$ mode,
which also contains the non-universal bare value $g$, determined by the equation
(\ref{qg}). From the equations (\ref{Ren}) we obtain
\begin{align}
Z_{x} & =Z_{z}=Z,  \notag \\
h_{R} & =Z^{2}h\left[ 1+\frac{1}{2}(\langle\pi_{z}^{2}\rangle+\langle
\pi_{x}^{2}\rangle)\right],  \notag \\
\rho_{R}f_{R} & =\rho_{r}Z^{2}f_{r}\left[ 1-\langle\pi_{z}^{2}\rangle\right], \notag \\
\rho_{xy,R}g_{R} & =\rho_{xy,r}Z^{2}g \left[ 1-\langle\pi_{x}^{2}\rangle\right].
\end{align}
Finally, $Z$ is fixed by the condition, that $\pi_{y}$ renormalizes the same
way, as $\pi_{x},$ which is due to 90$^{\circ}$ rotation symmetry in the
plane. This implies $h_{R}=Zh,$ such that
\begin{align}
Z & =1-\frac{1}{2}\left(
\langle\pi_{z}^{2}\rangle+\langle\pi_{x}^{2}\rangle\right),  \notag \\
\rho_{R} & \approx \rho_{r} \left[1
-\langle\pi_{x}^{2}\rangle\right],  \notag \\
f_{R} & \approx f_{r}\left[ 1
-2\langle\pi_{z}^{2}\rangle\right],  \notag \\
\rho_{xy,R}/\rho_{xy} & =1-\langle\pi_{z}^{2}\rangle,  \notag \\
\gamma_{R}/\gamma & =g_R/g= 1-2\langle\pi_{x}^{2}\rangle,   \label{Pert}
\end{align}
where we have introduced $\gamma=(\rho_{\parallel}-\rho_{\perp})/(2\rho
_{xy})$ and neglected small anisotropy terms in the second and third lines. Being rewritten through these quantities, the effective Lagrangian
reads
\begin{align}
L_{R}[\mathbf{n}] & =\frac{1}{2}\int d^{2}\mathbf{x}\left\{ \rho_{R}\left[
\mathbf{(\nabla}\pi_{z}\mathbf{)}^{2}+f_{r}\pi_{z}^{2}\right] +\rho _{xy,R}
\left[ (\nabla\pi_{x})^{2}+g_{R}+\gamma_{R}(\partial_{x}\pi_{x})^{2}-
\gamma_{R}(\partial_{y}\pi_{x})^{2}\right] \right.  \notag \\
& \left. +\frac{\rho_{xy,R}}{1-\pi^{2}}(\pi\nabla\pi)^{2}+\frac{\rho
_{xy,R}\gamma_{R}}{1-\pi^{2}}\left[ (\pi\partial_{x}\pi)^{2}-(\pi\partial
_{y}\pi)^{2}\right] \right\}.
\end{align}

\subsection{Renormalization group}
To perform RG analysis we introduce sharp momentum cutoff at scale $\Lambda$ and
vary $\Lambda$ from $\Lambda_{\mathrm{uv}}$ to the smallest possible scale;
in the following we replace accordingly the index $R$ at the renormalized
quantities by $\Lambda,$ denoting explicitly, at which infrared scale
they are evaluated. We also assume in the following that $f>g>\alpha$.
According to the obtained expressions, we perform
renormalization group procedure in two steps. At the first step we integrate
degrees of freedom at momenta scales $f_{\Lambda}^{1/2}<\Lambda<\Lambda_{\mathrm{uv}}.
$ In this range we can neglect small difference between $x-$ and $z-$ modes
in equations (\ref{Pert}) and obtain the standard flow equations of the $O(3)$ non-linear sigma model with small
easy-plane anisotropy
\begin{align}
\frac{dt_{\Lambda}}{d\ln(1/\Lambda)} & =t_{\Lambda}^{2}+t_{\Lambda}^{3},
\notag \\
\frac{d\ln Z_{\Lambda}}{d\ln(1/\Lambda)} & =-t_{\Lambda},  \notag \\
\frac{d\ln g_{\Lambda}}{d\ln(1/\Lambda)} & =\frac{d\ln \gamma_{\Lambda}}{
d\ln(1/\Lambda)}=\frac{d\ln f_{\Lambda}}{d\ln(1/\Lambda)}=-2t_ \Lambda,   \label{RG1}
\end{align}
where $t_{\Lambda}=T/(2\pi\rho_{\Lambda})$. In the first Eq. of (\ref{RG1})
we have added the two-loop term of the $O(3)$ model. The solution of Eqs. (\ref{RG1}) is well known,
\begin{align}
t_{\Lambda} & =\frac{t_{r}}{1-t_{r}\ln\left (\frac{\Lambda_{\mathrm{uv}}t_{\Lambda}
}{\Lambda t_{r}}\right )}, \notag \\
Z_{\Lambda} & =\frac{t_{r}}{t_{\Lambda}}=1-t_{r}\ln\left (\frac{\Lambda _{\mathrm{
uv}}t_{\Lambda }}{\Lambda t_{r}}\right ), \notag \\
\frac{g_{\Lambda}}{g_{r}} & =\frac{\gamma_{\Lambda}}{\gamma}=\frac {
f_{\Lambda}}{f_{r}}=\left( \frac{t_{r}}{t_{\Lambda}}\right) ^{2}=\left[
1-t_{r}\ln \left(\frac{\Lambda_{\mathrm{uv}}t_{\Lambda}}{\Lambda t_{r}}
\right)\right] ^{2},
\end{align}
where we have introduced $t_{r}=T/(2\pi\rho_{r})$. The scaling is stoped at $\Lambda=\Lambda_{f}$,
which fulfills
$f_{\Lambda_{f}}=\Lambda_{f}^{2}.$ The condition $t_{\Lambda_{f}}=1$ determines the
Kosterlitz-Thouless temperature in the absence of the in-plane anisotropy. At the
scales $g_{\Lambda}^{1/2}<\Lambda<f_{\Lambda}^{1/2}$ the $z$-mode is fully gaped, and we
obtain behavior of the coupling constants
\begin{align}
\frac{dt_{xy,\Lambda}}{d\ln(1/\Lambda)} & = \frac{d\ln f_{\Lambda}}{d\ln(1/\Lambda)} =0, \notag \\
\frac{d\ln Z_{\Lambda}}{d\ln(1/\Lambda)} & =-t_{xy,\Lambda}/2,  \notag \\
\frac{d\ln g_{\Lambda}}{d\ln(1/\Lambda)} & =\frac{d\ln \gamma_{\Lambda}}{d\ln(1/\Lambda)}=-2t_{xy,\Lambda},   \label{RG2}
\end{align}
which is in the XY universality class. The consideration of this regime is similar
to the case of quasi-two-dimensional easy plane model, \cite{Katanin} and yields the result for the Neel temperature in
Eq. (\ref{tc}) of the main text.


\begin{thebibliography}{99}
\bibitem{SM} All parameters of the model Hamiltonian are available upon request.
\end{thebibliography}
\end{document}